\documentclass[structabstract]{aa}  
\def \XMM {$XMM$-$Newton$\ }
\def \rchisq {$\chi_{\nu} ^{2}$}

\def\1p5{1.5DF}

\usepackage{txfonts}

\usepackage{latexsym,amsmath,amssymb}
\usepackage{natbib}
\usepackage{rotating}
\bibpunct{(}{)}{;}{a}{}{,}

\def \hcm {\hbox {\ifmmode $ atom cm$^{-2}\else atom cm$^{-2}$\fi}}

\def \chisq {$\chi ^{2}$}
\def \rchisq {$\chi_{\nu} ^{2}$}
\def\approxgt{\mathrel{\hbox{\rlap{\lower.55ex \hbox {$\sim$}}
        \kern-.3em \raise.4ex \hbox{$>$}}}}
\def\approxlt{\mathrel{\hbox{\rlap{\lower.55ex \hbox {$\sim$}}
        \kern-.3em \raise.4ex \hbox{$<$}}}}

\begin{document}

\title{X-ray variations in the inner accretion flow of Dwarf Novae} 
\author{\c{S}. Balman \inst{1} \thanks{\it email:solen@astroa.physics.metu.edu.tr} \and M. Revnivtsev\inst{2}}
\institute{Middle East Technical University, Dept. of Physics, In\"on\"u Bulvar{\i}, Ankara, Turkey, 06531
\and Space Research Institute, Russian Academy of Sciences, Profsoyuznaya 84/32, 117997 Moscow, Russia
\\
}
\date{}
\authorrunning{Balman \& Revnivtsev}
\titlerunning{X-ray variations in the accretion flow of Dwarf Novae}


\abstract
   {}
{Study of the inner disk structure of dwarf novae 
(i.e., nonmagnetic cataclysmic variables).}
{Power spectral analysis of the X-ray light curves obtained using the Rossi X-ray Timing Explorer ($RXTE$)
and  X-ray Multi-mirror Mission (\XMM) data. We fit such power spectra with a simple model that describes variability due to 
matter fluctuations. In addition, we perform cross-correlation analysis of simultaneous
UV and X-ray light curves using the \XMM\ data in order to determine time lags
between the different wavelength data.}
{
We show for five DN systems, SS Cyg, VW Hyi, RU Peg, WW Cet and T Leo that the UV and X-ray power
spectra of their time variable light curves are similar in quiescence. All of them show a break in their power spectra, which 
in the framework of the model of propagating fluctuations indicates inner disk truncation.
We derive the inner disk radii for these systems
in a range (10-3)$\times$10$^{9}$ cm. We analyze the $RXTE$ data of SS Cyg in outburst 
and compare it with the power spectra, obtained during the period of quiescence. We show that during the 
outburst the disk moves towards the white dwarf and recedes
as the outburst declines. We calculate the correlation  between
the simultaneous UV and X-ray light curves of the five DN studied in this work, using the \XMM data
obtained in the quiescence and find X-ray time lags of  96-181 sec. This
can be explained by the travel time of matter from a truncated
inner disk to the white dwarf surface.}
{
 We suggest that, in general, DN
may have truncated accretion disks in quiescence which
can also explain the UV and X-ray delays in the outburst stage and that
the accretion may occur through  coronal flows in the disk (e.g., rotating accretion disk coronae).
Within a framework of the model of propagating fluctuations the comparison of the X-ray/UV time lags observed by us in the 
case of DN systems with those, detected for a magnetic Intermediate Polar allows us to make a rough estimate of the viscosity 
parameter $\alpha\sim0.25$ in the innermost parts of the accretion flow of DN systems.}
\keywords{
X-rays: stars,binaries --- accretion,accretion dics --- 
binaries: close --- stars: dwarf novae --- cataclysmic variables
 --- stars: individual : SS Cyg, RU Peg, VW Hyi, WW Cet, T Leo
}

\maketitle


\section{Introduction}

Cataclysmic variables (CVs) are interacting binaries
hosting a white dwarf (WD) primary star accreting material from a
late-type main sequence (MS) star. Accreting material forms a disk 
that is expected to reach all the way to the WD
in cases where the magnetic field of the WD is weak ( $B$ $<$ 0.01
MG) and such systems are referred as nonmagnetic CVs \cite[see][for a review]{warner95} 

In our work we consider dwarf novae -- a subclass of non magnetic cataclysmic variables.
In such systems the matter, which is transferred from the secondary star to the 
Roche lobe of the primary, does not form a stationary flow to the WD surface, but 
the mass transfer rate is diminishing towards the WD during the so called quiescent state.
These states are interrupted every few weeks to tens of years by the enhanced accretion flow
(outbursts) that lasts days to weeks and the systems significantly brightens (bolometrically).

The material in the inner disk of dwarf novae initially moving with the Keplerian velocity 
dissipates its kinetic energy in order to accrete onto the slowly rotating WD creating a boundary layer \citep{lyndenbell74,kippenhahn78,narayan93,godon95}.

Observations show that during the quiescence (low-mass accretion rates in the innermost parts of the flow) a significant or 
dominant fraction of the bolometric emission of dwarf novae is emitted via the bremsstrahlung process
from an optically thin hot plasma in 
the hard X-rays \citep{patterson85,narayan93}. 
Typical characteristic of the quiescent X-ray emission
is a multi-temperature  quasi-isobaric 
cooling flow model of plasma emission with temperatures of 6-50 keV 
and an accretion rate of 10$^{-12}$-10$^{-10}$ M$_{\odot}$/yr with
$L_x<10^{33}$ erg s$^{-1}$ \citep{perna03,pandel05,kuulkers06,rana06,balman11}.

Observational appearance of quiescent dwarf novae indicates that significant part of the energy release during ongoing 
accretion is happening in the optically thin region near the white dwarf. Thus the accretion flow is likely changing 
its character from optically thick disk-like flow in the its outer parts to an optically thin "corona"-like flow close to the WD.

Observationally, such a truncation of the optically thick accretion disk in dwarf novae in 
quiescence was invoked due to the observed time lags between the optical and UV fluxes at the 
rise phase  of the outbursts \cite[see reviews][]{lasota01,lasota04}, or due to unusual shape of 
the optical spectra or light curves of DNe \cite[see e.g.][]{linnell05,kuulkers11}.  Theoretical support for such two-phase 
flow was given by a model of the disk evaporation of \cite{meyer94}.
This model was later elaborated to show that the disk evaporation (coronal "syphon" flow) may create 
optically thick-optically thin transition regions at various distances from the WD \citep{liu97,mineshige98}.

Attempts to obtain a map of the accretion disk in cataclysmic variables were done with the help of the eclipse mapping 
method \cite[e.g.][]{horne85}, but this method has its own limitations and it is not very sensitive to the innermost parts of 
the accretion disk.

Recently, an additional diagnostic tool was proposed -  the aperiodic variability of brightness of sources.
Virtually all accreting sources demonstrate aperiodic variability of their brightness over a 
wide range of time scales (see \citealt{bruch92} for a review of aperiodic variability of cataclysmic variables). While 
the long time scale variability might be created at the outer parts of the accretion disk 
\cite[see e.g.][]{warner71}, the relatively fast time variability (at $f>$few mHz) originates in the inner parts of 
the accretion flow \cite[see e.g.][]{bruch00,baptista04}

Properties of this noise is similar to that of the X-ray binaries with neutron stars and black holes.
These properties are quite peculiar, which does not allow to explain it as a sum of 
independent burst-like events in the region of main energy release. In particular, the 
variability demonstrates very tight relation between the rms amplitude of variations and the average flux 
level \cite[see e.g.][]{uttley01,scaringi12}. The variability spans over an extremely wide range of 
frequencies with the same power-law slopes \cite[see e.g.][]{churazov01}. Now, the widely accepted 
model of origin of this aperiodic flicker noise is a model of propagating fluctuations 
\citep{lyubarskii97,churazov01,uttley01,arevalo06,revnivtsev09,revnivtsev10,uttley11}. 

In the framework of this model, the modulations of the light are created by variations of the instantaneous value of 
the mass accretion rate in the region of the energy release. These variations of the mass accretion rate, in turn, 
are inserted into the flow at all radii of the accretion disk due to the stochastic nature of its viscosity and then 
transferred toward the compact object. Fast variations of the mass accretion rate inserted into the flow at 
distances closer to the central object modulate the mass flow incoming into these regions from outer parts of the 
accretion disk. 

This model predicts that the truncated accretion disk should lack some part of its variability at high Fourier 
frequencies, i.e. at the time scales shorter than typical time scale of variability at the inner edge of the disk. 
This prediction was checked for the accreting systems, in which the disks are indeed truncated due to the interaction 
with the compact object magnetospheres, in particular - accreting magnetic neutron stars \citep{revnivtsev09} and 
accreting magnetic white dwarfs \citep{revnivtsev10}. The revealed breaks in the power spectra of these accreting 
binaries allowed one to make estimates of the inner radius of the accretion disk. In particular, in the work of 
\cite{revnivtsev11} it was shown that the inner truncation radius of the accretion disk in EX Hya estimated from the 
variability arguments was quite compatible with those estimated with the help of completely different physical effects.

In this paper we would like to apply the similar diagnostic tool to make estimates of the inner boundary of the 
optically thick accretion disk in non-magnetic WDs -- dwarf novae. We use \XMM and $RXTE$ data to study the broad-band 
noise in DN and  calculate the inner disk radii for five systems, SS Cyg, VW Hyi, RU Peg, WW Cet, and T Leo. 
Basing on the non-detection of any periodic X-ray light variations (with the very tight upper limit) in flux of the enlisted 
systems we assume that they contain non-magnetic WDs, while such a classification for SS Cyg is challenged by some 
authors \cite[e.g.][]{lombardi87,giovannelli99}. This is not particularly important for our study because we study the 
accretion disk truncation and the origin of this truncation might have different nature in different sources.

We infer from our results that these systems have truncated disks at large radii
and that some form of rotating coronal flow (e.g., ADAF disks and/or accretion disk coronae)
exits in DN systems where the material is transported to the surface of the WD. 
  
\section{The Data and Observation}
 
The \XMM Observatory \citep{jansen01} has three 1500 cm$^2$ 
X-ray telescopes each with an European Photon Imaging Camera (EPIC) 
at the focus; two of which have Multi-Object Spectrometer (MOS) CCDs
\citep{turner01} and the last one uses pn CCDs \citep{struder01}
for data recording.
The Optical Monitor (OM), a photon counting instrument,
is a co-aligned 30-cm optical/UV telescope, providing for the first time the
possibility to observe simultaneously in the X-ray and optical/UV regime from a single platform 
\citep{mason01}.
For our timing purposes we utilized the data collected with the
EPIC pn cameras in the partial or full window imaging mode,
and the Optical Monitor OM data using the fast imaging mode ($\ge$0.5 sec time
resolution)  with the UVW1 filter (240-340 nm). The time resolution of
imaging modes are 70 ms, the timing mode observations have 30 $\mu$s resolution
and the burst mode goes down to 7 $\mu$s in time resolution.
We summarize the archival \XMM data used in this analysis on Table 1. We utilized the EPIC pn data
and the MOS data, when necessary, for our analysis since EPIC has the adequate
sensitivity and timing resolution for our study. 
A medium optical blocking filter was used with all the EPIC cameras. 
We analysed the pipeline-processed data using Science Analysis Software (SAS)
version 11.0.0. Data
(single- and double-pixel events, i.e., patterns 0--4 with Flag=0 option) were extracted from
a circular region of radius 30$^{\prime\prime}$ for pn, (MOS1 and MOS2)
in order to perform temporal analysis together with the background events extracted from a
source free zone normalised to the source extraction area.
We checked/cleaned the pipeline-processed events file from  the existing flaring 
episodes during the analysis for all the objects. The OM data was analyzed using the {\sc omfchain}
with 1 sec time resolution.

We used Rossi X-ray Timing Explorer ($RXTE$; \citealt{bradt93}) archival data for 
the analysis of the outburst data of SS Cyg  and the quiescence data for comparison.
The data were obtained by the Proportional Counter Array (PCA; \citealt{jahoda06}) 
instrument aboard $RXTE$. The PCA consists of five xenon-filled detector units (PCUs) with a total effective area of 6200 cm$^2$ at 5 keV. 
It is sensitive in the range 2-60 keV, the energy resolution is 17$\%$ at 5 keV, and the time resolution capability is 1 $\mu$s .
A log of observations can be found in Table 1. RXTE/PCA background was estimated with the help of the
model appropriate for faint sources. Light curves of the source
were extracted using the standard procedures (i.e., {\sc saextrct}) from the data of 1-3 PCUs mainly in the 
entire PCA energy band
utilizing the "Standard 1" data with 0.125 sec time resolution. As an 
additional check of the results we have analysed the Good Xenon mode data 
in the energy band 3-20 keV and obtained similar results. All event arrival 
times were corrected to the solar system barycenter.

All light curves were background 
subtracted for the analysis. 
The data and light curves were analysed using HEASOFT version V6.9 . 

\begin{table*}
\label{1}
\caption{The log of observations (list of Observation IDs) used for the analysis in our work. All $RXTE$ data is of SS Cyg.}
\begin{tabular}{ll}
\hline
\hline
\\

$RXTE$  &
20033013100--20033014200, 50011018000--50011019700, 50012010101--50012010109
\\
Quiescence &
\\
$RXTE$ & 
40012010200, 40012010500, 40012010600, 50011013100--50011014300, 95421010100, 95421010200, 
\\
X-ray dips &
95421010800, 95421011000, 954210200--954210204
   \\
$RXTE$  &   
40012011100--40012011800, 50011015800--50011017100, 95421010300--95421010301, 
\\
X-ray Peak & 50012010104--50012010108
\\
\XMM & 0111310201 (SS Cyg), 
0111970301 (VW Hyi),  0111970901 (WW Cet),  0551920101 (RU Peg), 
\\
Quiescence &  
0111970701 (T Leo) 
\\

\hline
\end{tabular}
\end{table*}

\section{Analysis and Results}
\subsection{SS Cyg}

We prepared light curves using the standard procedures as described in the previous section. 
In order to measure the broad-band noise, we derived and averaged several power spectra for each source 
from the calculated light curves.
The power spectral densities (PDS) expressed were calculated in terms of the fractional rms amplitude squared
following from \cite{miyamoto91},
$$ P_j=2|A_j|^2/N_{ph}C\ \ \ \ \ \  
A_j=\sum x_n e^{i \omega_n t_n} $$. In this prescription, $t_n$ is the time label for each time bin, $x_n$ is the number of counts in
these bins, $N_{ph}$ is the total number of photons in each light curve and C is the average count rate in each time
segment used to construct PSD.  The white noise levels were subtracted hence
leaving us with the rms fractional variability of the time series in units of $(\rm{rms}/\rm{mean})^2$/Hz.
Next, we multiplied the rms fractional variability per hertz with the frequencies at which they were 
calculated
yielding an rms fractional variability squared, thus our PDS are ${\nu} P_{\nu}$ versus ${\nu}$. 
For the model fitting we used a simple analytical model 

$$ P({\nu}) \propto {\nu}^{-1} \left(1 + \left(\frac{\nu}{\nu_0}\right)^4 \right)^{-1/4} $$


\noindent
which was proposed to describe the power spectra of sources with truncated 
accretion disks \cite[see e.g.][]{revnivtsev10,revnivtsev11}. 

We used the archival $RXTE$ data of SS Cyg in quiescence and outburst listed on Table 1 to derive the  
broad-band noise of the source in different states (i.e., accretion rates). For the outburst phases,
we investigated times during the X-ray suppression (e.g., the X-ray dips; optical peak phases of the outburst), 
and the X-ray peak.
This, in turn is expected to show the motion of the
flow in the inner regions of the disk and the geometry of the inner disk. 
For details of the source spectra
and light curves during outburst and quiescence, \cite[see][]{mcgowan04}.
Figure 1 shows a compilation of the 
PDS of SS Cyg obtained from $RXTE$ data. This includes the $RXTE$ PDS of the quiescent phase data,
the $RXTE$ PDS during the optical bright phase (i.e., X-ray dips) and the PDS 
during the X-ray peak of the outburst. 

The source PDS from the $RXTE$ data are fitted with the described model in the above paragraph;  
${\nu} P_{\nu}$=$\rm{P2}\times$$\left(1 + \left(\frac{\nu}{\rm{P1}}\right)^4 \right)^{-1/4}$ that uses two parameters 
P1 being the break frequency noted as $\nu_0$ in the analytical model and the other is P2, the normalization that
determines the power level of the flat part of the broken power-law.
The reduced \chisq values of the fits are 0.62, 1.5, and 0.4 for the 
quiescence, the X-ray peak and the X-ray dips,
respectively.
The resulting break frequencies are displayed in Table 2 and the 
corresponding disk truncation radii are 
calculted and displayed in the same Table 
using the simple relation $\nu_0$ $\simeq$ (GM$_{WD}$/R$_{in}^3$)$^{1/2}$/2$\pi$  (assuming\ 1 M$_{\odot}$ for the WD mass).
The broad-band noise structure of the Keplerian disks often show $\propto$ $f^{-1\dots-1.3}$ dependence on frequency \citep{churazov01,gilfanov05} and this
noise will show a break if the optically thick disk truncates as the Keplerian motion subsides. We detect that all $RXTE$ 
PDS show breaks and this implies that the disk in SS Cyg is truncated at large distances compared
with the size of the WD.

As clearly shown in Table 2, we detect that the inner edge of the disk is further out during the quiescence and the
disk moves inwards to about 1.1$\times 10^9$ cm during the peak of the optical outburst and starts moving out to 
about 3.3$\times 10^9$ cm
during the X-ray peak of outburst and finally it reaches to about 5.5$\times 10^9$ cm at the quiescent flux levels 
after the outburst has subsided.

The quiescent \XMM\ data of SS Cyg obtained at a different date is used to validate the 
quiescent $RXTE$ PDS and check the disk
truncation radius.  The two fitted quiescent PDS are very similar as displayed in Figure 2. 
The resulting  break frequency and the disk truncation radius are given in Table 2.
The reduced \chisq of the fit using the \XMM\ (EPIC pn)  PDS is 1.4\ .
In addition, we 
have calculated the  PDS in the UV from
the OM data (\XMM) and included it in Figure 3 for comparison.
The break frequency and the disk truncation radius 
calculated from the UV overlap with the X-ray results within the 95$\%$ confidence level error ranges. 
The reduced \chisq of the fit to the \XMM\ (OM) PDS is 0.4\ .

\begin{figure*}
\centerline{
\includegraphics[width=6cm,height=8cm,angle=270]{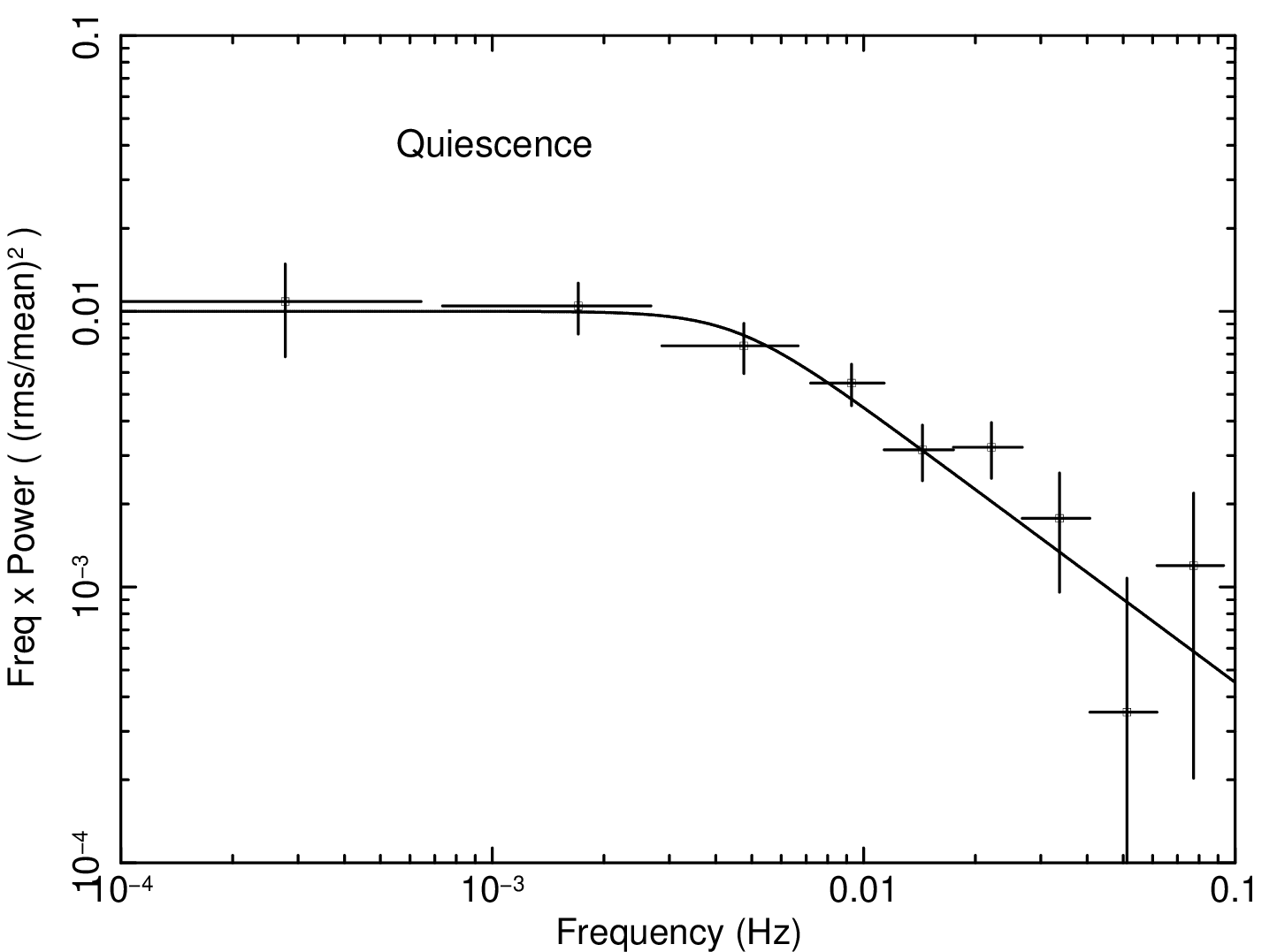}
\includegraphics[width=6cm,height=8cm,angle=270]{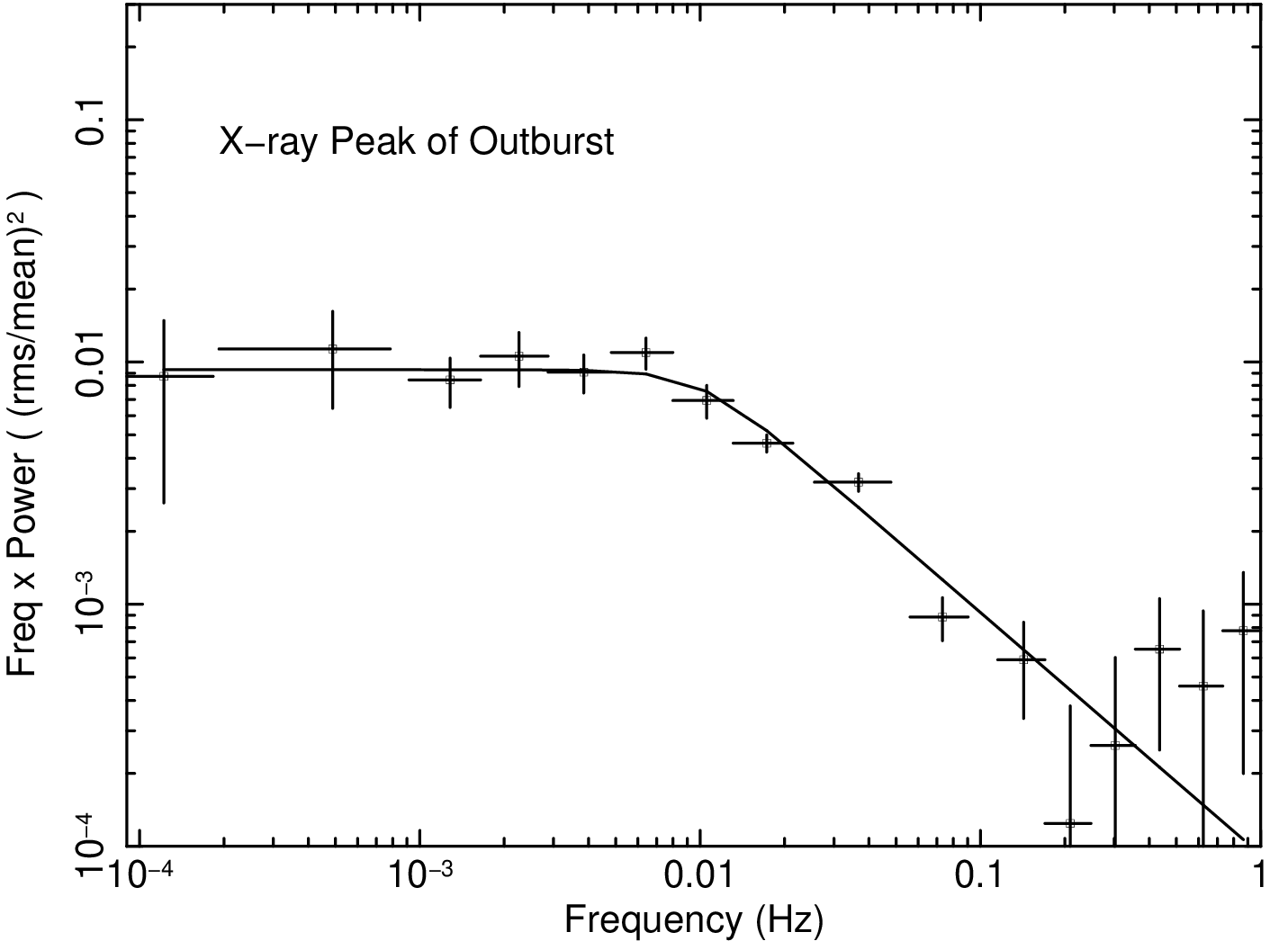}}
\centerline{
\includegraphics[width=6cm,height=8cm,angle=270]{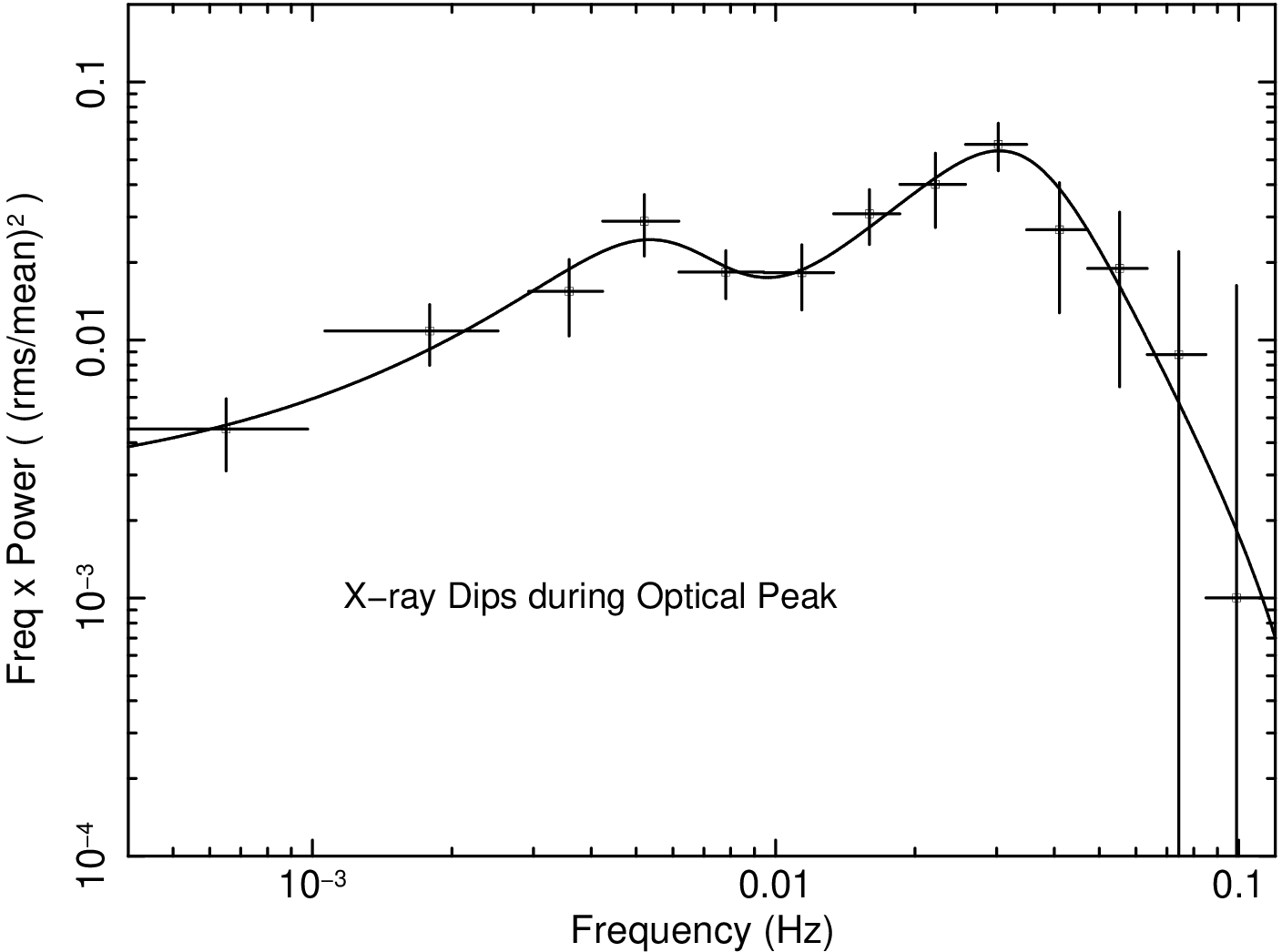}
\includegraphics[width=6cm,height=8cm,angle=270]{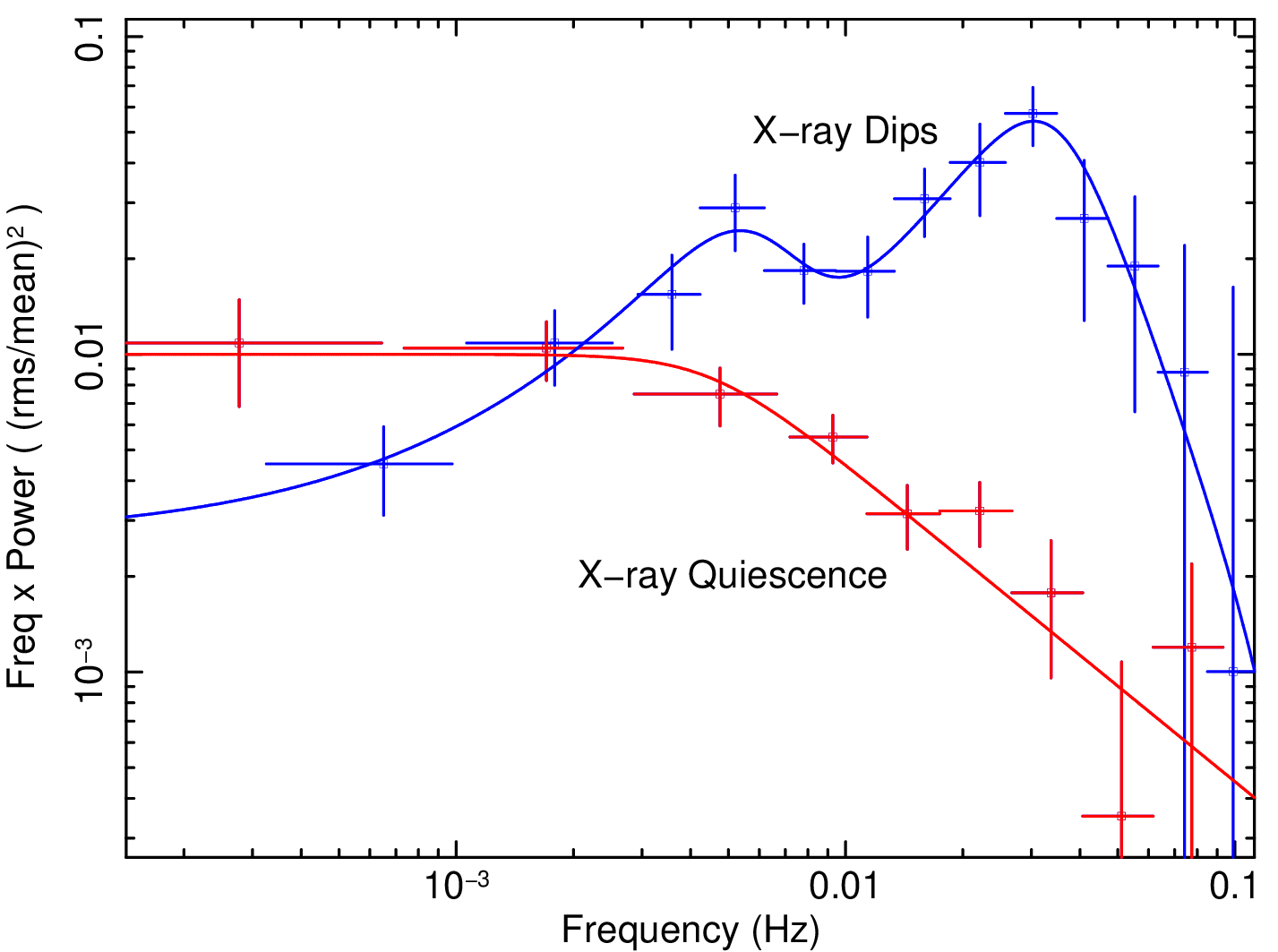}}
\caption{Power Spectra of SS Cyg in outburst obtained from a collection of archival $RXTE$ data listed on Table 1. The PDS at
different
times are ordered as the quiesence on the top left hand side, the X-ray peak on the top right hand side, and the X-ray dips during
the optical peak on the bottom left hand side. On the bottom right hand side, the PDS of SS Cyg in quiescence and during the
X-ray suppression (optical peak) is shown for comparison.
The solid lines show the fit with the propagating fluctuations model for the top figures and for the PDS of the X-ray dips
two Lorentzians
along with the propagating fluctuations  model are used to achive the best fitting 
results.  The reduced \chisq values of the fits are 0.62, 1.5, and 0.4 
for the quiescence, the X-ray peak and the X-ray dips,
respectively.}
\end{figure*}

\subsection{Other DNe}

In order to check if the truncation of the inner disk radius is unique to SS Cyg, we used 
the archival \XMM\ data of other different kinds of
dwarf novae. We prepared light curves in the same manner and prepared PDS from 
simultaneous X-ray (EPIC) and UV (OM) data.
Figure 4-9 shows the PDS of four other dwarf nova; RU Peg (U Gem type), VW Hyi (SU UMa type), WW Cet (Z Cam type), and T Leo (SU UMa type). We omitted the UV 
PDS of WW Cet and T Leo due to low statistical quality. 
We applied the same fit (using the same functional form) 
to the PDS of these dwarf novae as was done for SS Cyg.
The fits to the various PDS yield reduced \chisq in a range 0.2-1.4 .
The particular values are indicated in the figure captions.
All PDS indicate disk 
truncation at radii larger than the size of the WD. 
We display the break frequencies calculated using the X-ray data and the 
disk truncation radii on Table 2. The inner disk radii are calculated as 
was done for SS Cyg and 
we assumed WD masses as listed by Pandel et al. (2005). 
The break frequency and the disk truncation radii
calculated from the UV overlap with the X-ray results within the 95$\%$ confidence level error ranges.

The orbital period of
VW Hyi was removed from the UV PDS by modeling a narrow Loretzian and this period is not detected
in the X-ray PDS. We underline that our UV data for all five DNe 
do not have long enough duration to derive the
PDS characteristics securely at low frequencies. Thus, we can not infer if there is
any contribution to red noise from other 
components of flickering noise like the accretion hot spot, overflowing accretion stream, and the spiral
waves as proposed using optical wavelength data \cite[see e.g.][]{bruch00,baptista04}.
 
\begin{figure}
\includegraphics[width=6cm,height=8cm,angle=270]{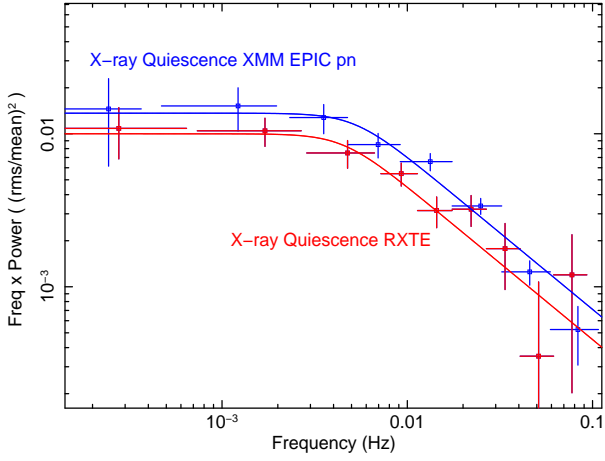}
\caption{X-ray power spectra of SS Cyg in quiescence obtained from the \XMM\ EPIC pn data.
Dark blue data points are of the \XMM\ EPIC pn PDS and the overlayed red data points are
of the $RXTE$ PDS. The solid lines
are the fitted curves using the propagating fluctuations model.
The reduced \chisq of the fit to the \XMM\ (EPIC pn) PDS is 1.4 .}
\end{figure}

\begin{figure}
\includegraphics[width=7cm,height=8cm,angle=270]{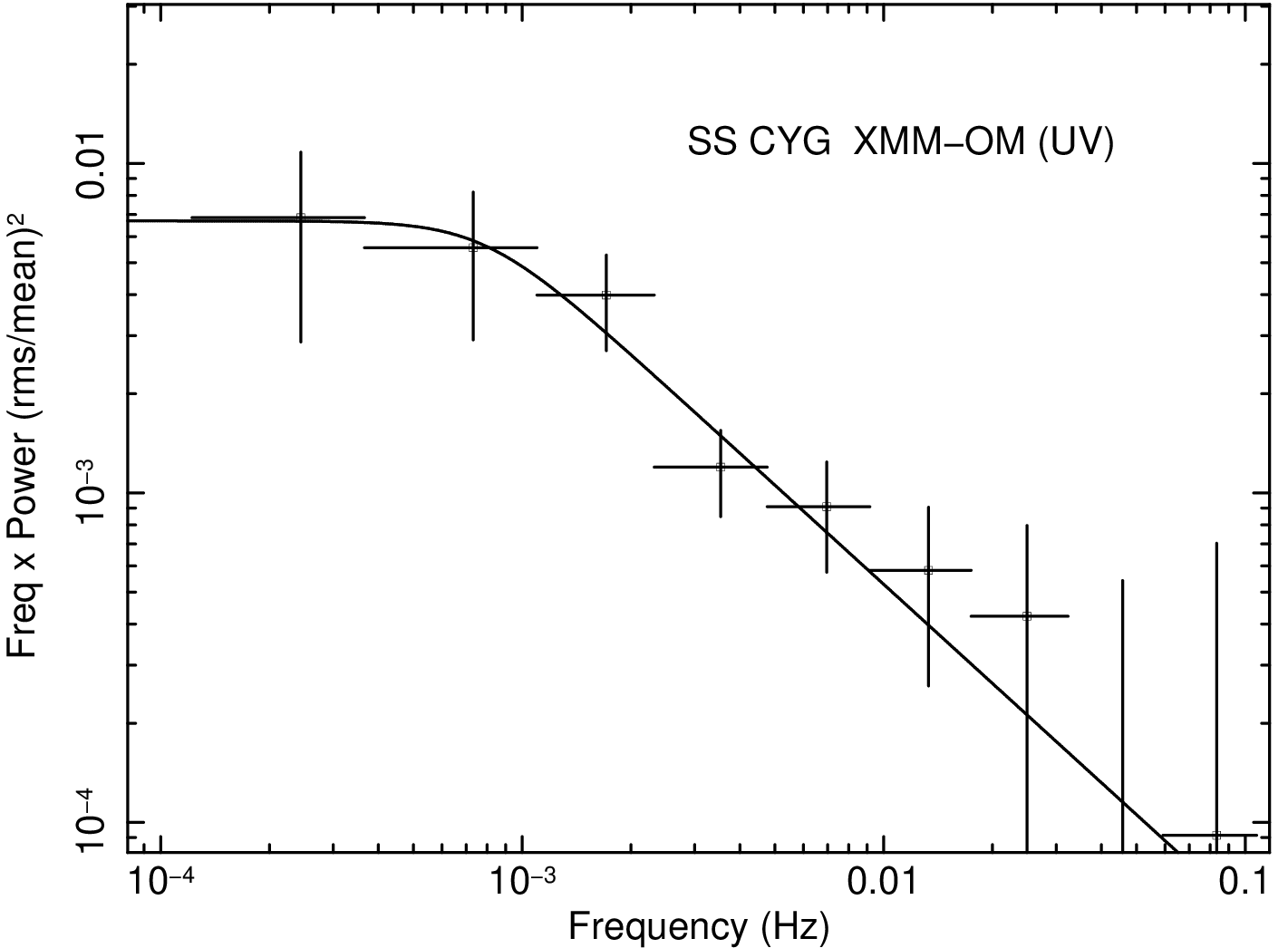}
\caption{UV power Spectrum of SS Cyg in quiescence obtained from the \XMM\ OM data. Black solid line
is the fitted curve using the propagating fluctuations model.
The reduced \chisq of the fit to the \XMM\ (OM) PDS is 0.4 .}
\end{figure}

\begin{figure}
\includegraphics[width=6cm,height=8cm,angle=270]{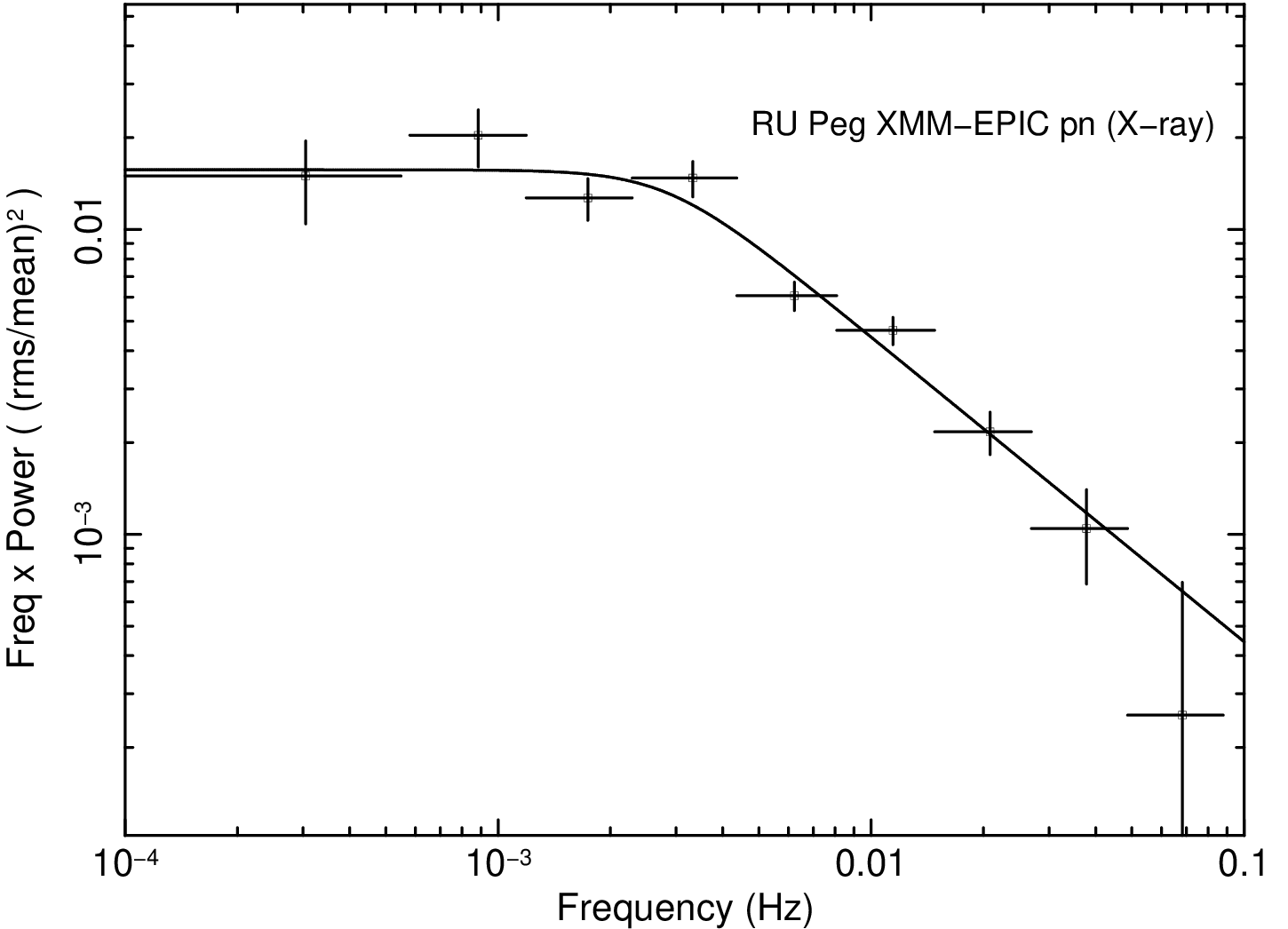}
\caption{X-ray power Spectrum of RU Peg in quiescence obtained from the \XMM\ EPIC pn data.
Black solid line
is the fitted curve using the propagating fluctuations model.
The reduced \chisq of the fit to the \XMM\ (EPIC pn) PDS is 1.5 .}
\end{figure}

\begin{figure}
\includegraphics[width=6cm,height=8cm,angle=270]{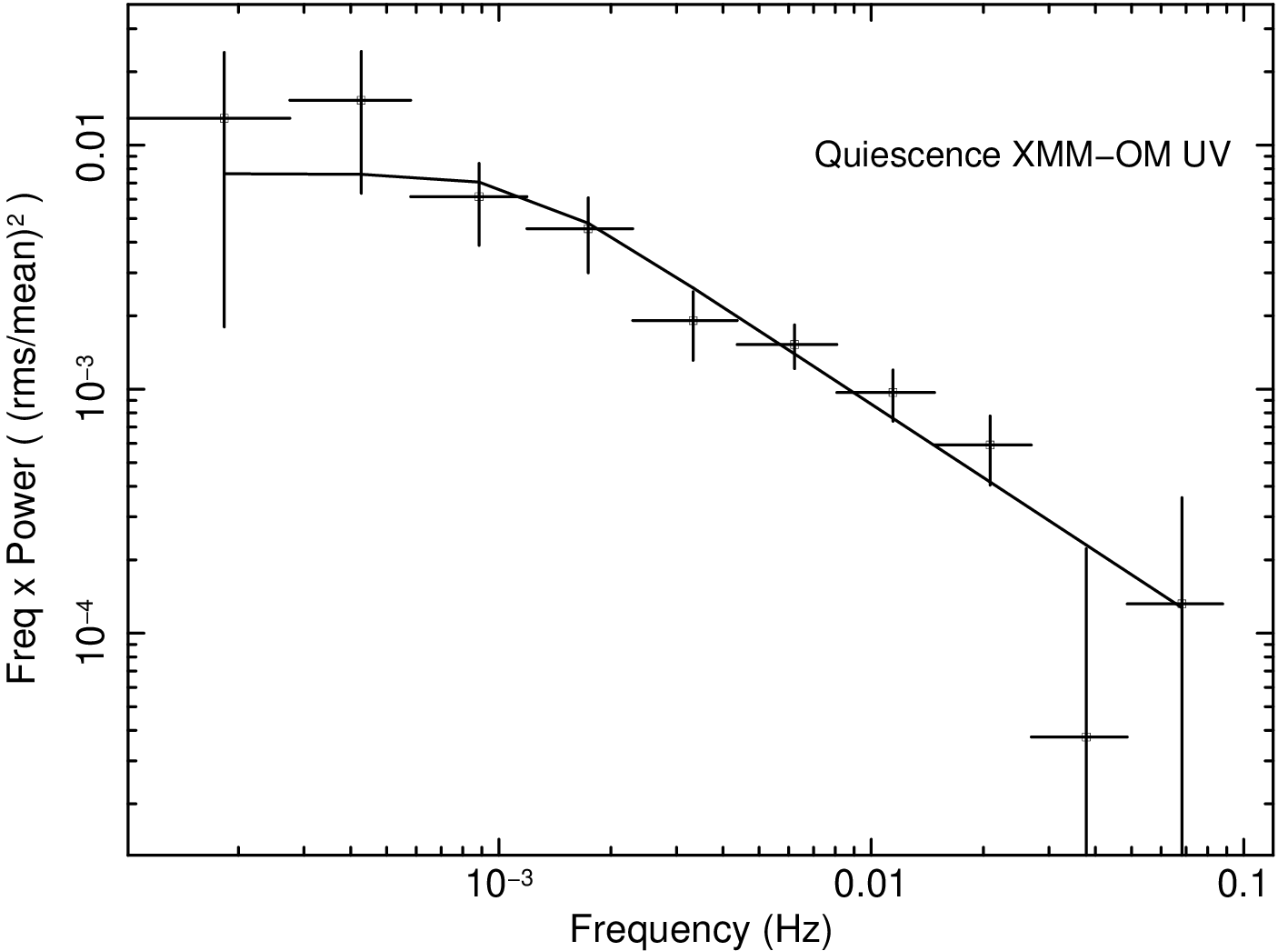}
\caption{UV power spectrum of RU Peg in quiescence obtained from the \XMM\ OM data.
Black solid line
is the fitted curve using the propagating fluctuations model.
The reduced \chisq of the fit to the \XMM\ (OM) PDS is 0.8 .}
\end{figure}

\begin{figure}
\includegraphics[width=6cm,height=8cm,angle=270]{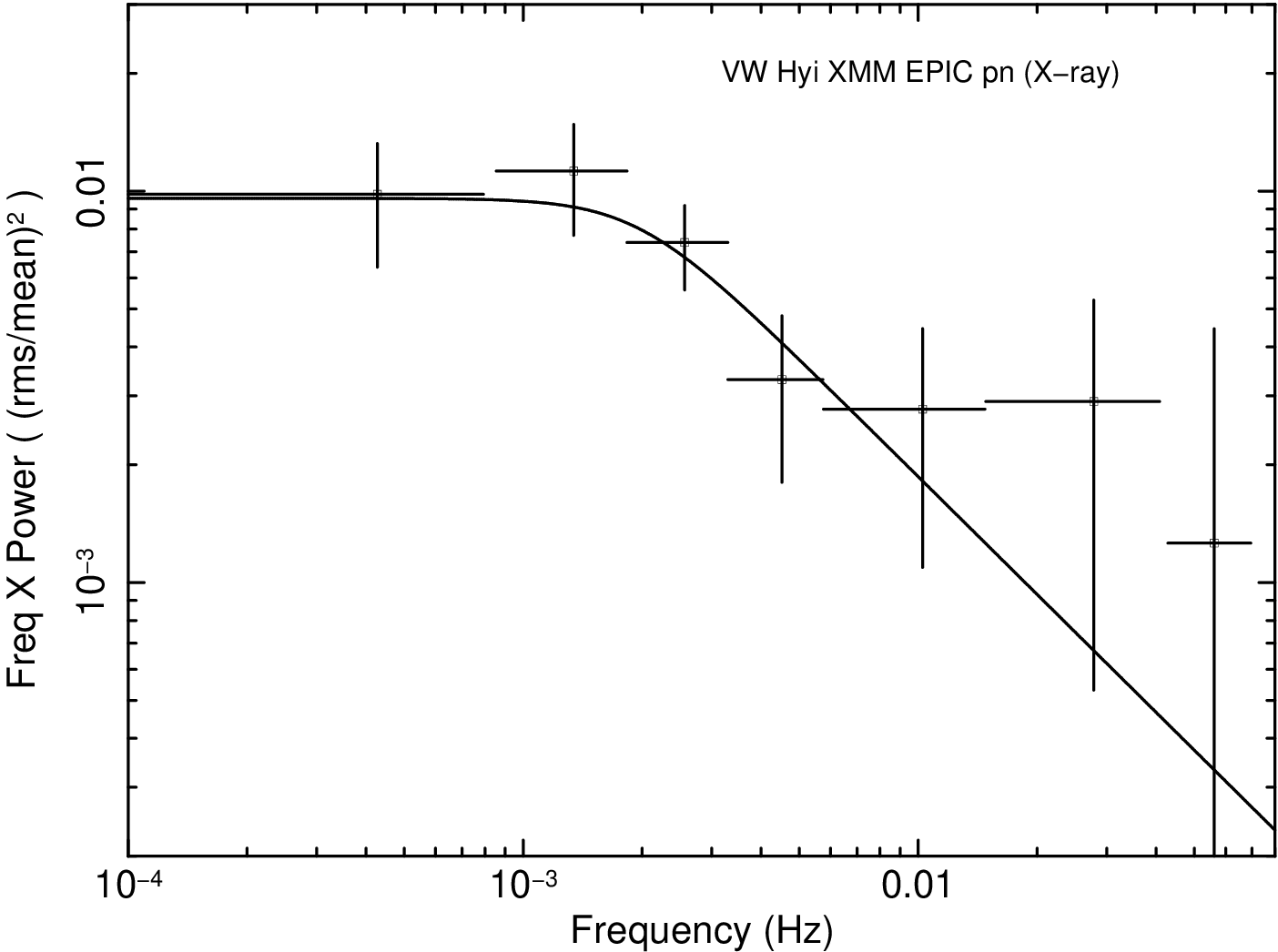}
\caption{X-ray power spectrum of VW Hyi in quiescence obtained from the \XMM\ EPIC data. 
Overplotted curve is the fitted  propagating fluctuations model.
The reduced \chisq of the fit to the \XMM\ (EPIC pn) PDS is 0.5 .}
\end{figure}

\begin{figure}
\includegraphics[width=6cm,height=8cm,angle=270]{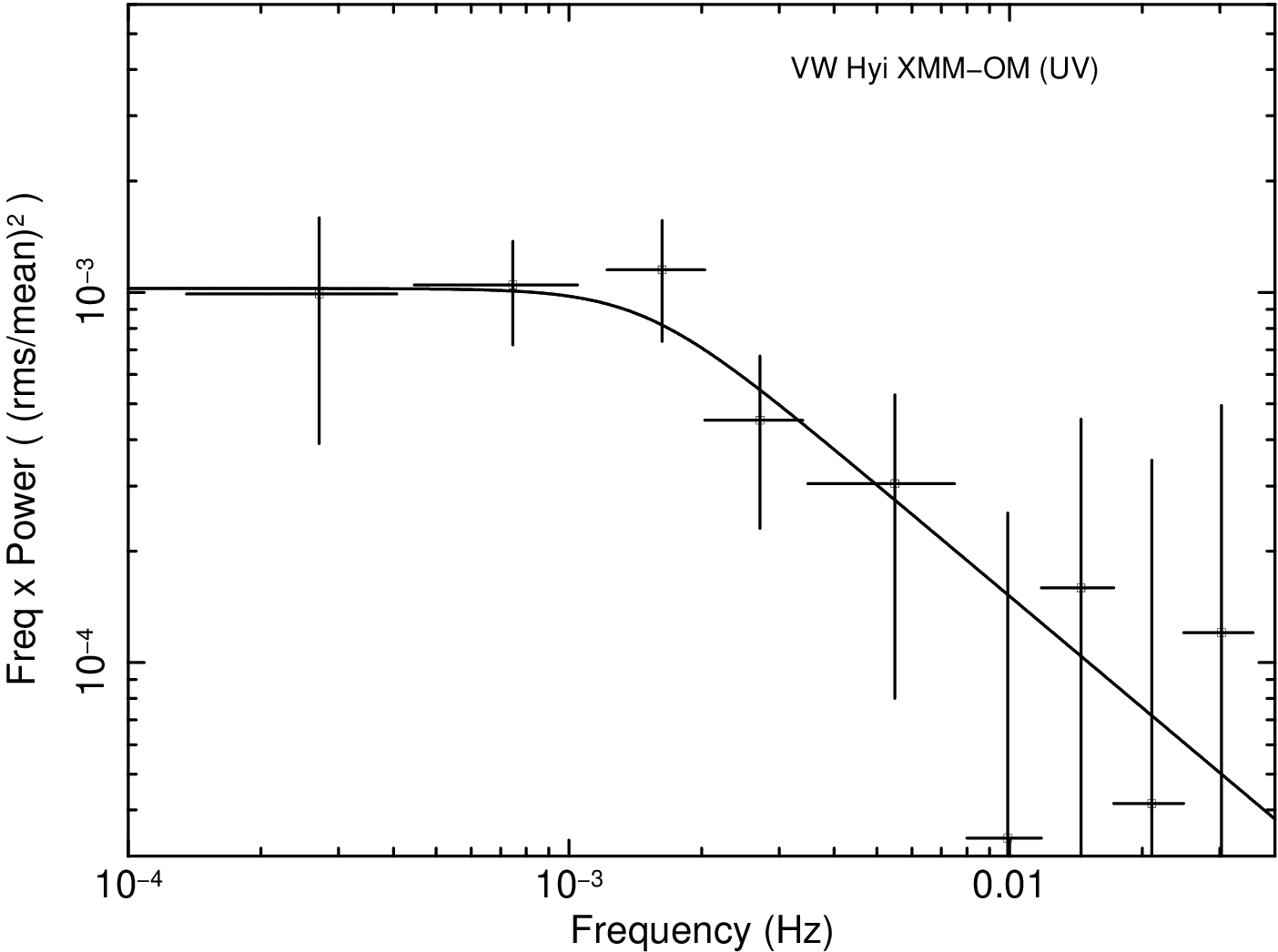}
\caption{UV power spectrum of VW Hyi in quiescence obtained from the \XMM\ OM data. 
Overplotted curve is the fitted propagating fluctuations model.
The reduced \chisq of the fit to the \XMM\ (OM) PDS is 0.4 .}
\end{figure}

\begin{figure}
\includegraphics[width=6cm,height=8cm,angle=270]{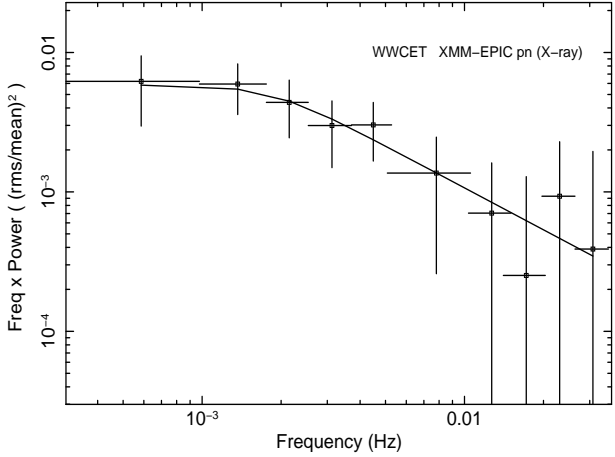}
\caption{X-ray power spectrum of WW Cet in quiescence obtained from the \XMM\ EPIC pn data. 
Overplotted is the fitted curve using the the propagating fluctuations model.
The reduced \chisq of the fit to the \XMM\ (EPIC pn) PDS is 0.2 .}
\end{figure}

\begin{figure}
\includegraphics[width=6cm,height=8.2cm,angle=270]{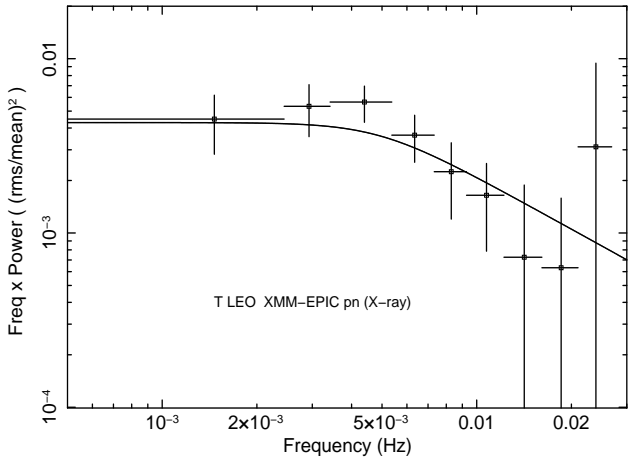}
\caption{Power spectrum of T Leo in quiescence obtained from the \XMM\ EPIC pn  data.
Overplotted is the fitted curve using the power relation for the noise
from the propagating fluctuations model.
The reduced \chisq of the fit to the \XMM\ (EPIC pn) PDS is 0.7 .}
\end{figure}

\subsection{The cross-correlations}

If the model of the propagating fluctuations correctly describes the time variability of 
the X-ray and optical/UV flux of DNe, then we should expect some particular way of correlations 
of brightness of systems at these energies. In particular, the optical/UV light, generated as  
reprocessing of the X-ray emission from the central regions should be lagged with respect to the 
X-rays consistent  with the light crossing time (typically less than a second for  majority of DNe, 
but $\sim$5-6 seconds for the largest binaries, like RU Peg). But the optical/UV light 
variations, generated as an energy release of variable mass accretion rate at the inner edge 
of the accretion disk, should lead the X-ray emission with the time lag equal to the 
time needed for the matter to travel from the inner edge of the disk to the 
central regions of the accretion flow in the vicinity of the WD, where the bulk of the
X-ray emission is generated.

In order to study this issue,  we calculated the cross-correlation between the two
simultaneous light curves (X-ray and UV), using the archival \XMM data utilizing {\sc HEAsoft} task {\sc crosscor}.
In order to obtain the CCFs (cross-correlation functions), we devided our datasets into several pieces using 1-5 sec bining in light curves
and calculated CCFs for each of them.

The CCF at each lag j, CCF(j) is computed with the expression:  

$$ CCF(j) = \frac{\sum{t}(x_{uv}(t)-\bar{x}_{uv})(x_{x-ray}(t+j)-\bar{x}_{x-ray})}{N(j)(\sigma_{uv}^2\sigma_{x-ray}^2)^{1/2}}$$

\noindent
where j=0,$\pm\Delta$t.... The summation goes from t=$\Delta$t to (N-j)$\Delta$t for zero and positive j
and from t=(1-j)$\Delta$t to N$\Delta$t for negative j. N is the total number of points in the light curve
while N(j) is the number of pairs that contribute to the calculation of CCF at lag j. The variances are
are source variances.
The  resulting correlation coefficients from the above expression
as a function of time lag is shown in Figure 10 for all the five dwarf nova, we analyzed.
We note that this correlation analysis was already presented for RU Peg in Balman et al. (2011)
and for VW Hyi in Pandel et al. (2003), but we include them here for comparison and completeness of 
our analysis.
The cross-correlation coefficient is normalized to have a maximum value of 1. 
The error bars are evaluated as root mean square deviations from the average value of 
the cross-corelation of each measurement (made by using the segments of the whole light curve).

The curves for all the dwarf novae show a clear asymmetry indicating that some part of 
the UV flux is leading the X-ray flux.
In addition, we detect a strong peak
near zero time lag for RU Peg, WW Cet and T Leo suggesting a significant zero-lag correlation between the X-rays
and the UV light curves. Our reanalysis of the cross-correlation for VW Hyi also reveals
an existence of correlation at zero time lag in addition to Pandel et al. (2003) 
which we include in Figure 10.
The positive time-lag leading to an  asymmetric profile for the four dwarf novae above 
and the shifted profile for SS Cyg show that the
X-ray variations are delayed relative to those in the UV.

In order to calculate an average time-lag that would produce the asymmetric or shifted
profile, we fitted the varying cross-correlation by two Lorentzians 
with time parameter fixed at 0.0 lag (not assumed for SS Cyg) and the other set as free.
A Lorentzian has
a functional form $F(y)$=$\rm{P3}\times$$\left(1 + \left(\frac{2. x-\rm{P1})}{P2}\right)^2 \right)^{-1}$
where P1 is the time parameter which yields the time lag, P2 is the FWHM of the Lorentzian
and P3 is the normalization. We report only the parameter P1 for the measure of the time lag and its erorr.
Our PDS are broken power laws as a result  our ACF/CCF can not be Gaussian-shaped.
In order for the ACF/CCF to be Gaussian, our PDS should also be Gaussian-shaped since the PDS is a Cosine Transfrom
of the ACF (auto correlation function). Our expected ACF/CCF should be more extended than a Gaussian
where a Lorentzian is, then, an acceptable choice.
The resulting fits
are displayed in Figure 11. The delay times vary in a range  70-240 sec for the five DN. 
The fit for RU Peg is taken from \cite{balman11}. The reduced \chisq of the fits in Figure 10
are 0.8, 0.4, 0.45, 1.2, and 0.45 from the top to the bottom panel of the figure, respectively.

In order to double check these generic fits, we constructed a more 
physically motivated plot of time-correlation using the shape of the 
auto-correlation of the X-ray light curves. Assuming 
that the zero-time lag signal on the original cross-correlation plot is 
created by a simple transformation of the X-ray 
variability into the UV light, we have subtracted that part adopting the shape of X-ray auto-correlation function. 
The amplitude of this zero-time lag component was taken to provide smooth behaviour of the subtracted cross-correlation 
function across zero delay. The subtracted cross-correlation 
functions directly yield the shifted time-lag
component. This is applied to the four DNe that show the zero-time lag 
in their 
cross-correlation and the residuals are fitted with a single Lorentzian to calculate the
time-lag. The resulting fitted residuals are displayed in Figure 11
and the measured time-lags are listed on Table 2 along with the reduced \chisq of the fits. Our adopted approach is simple, 
but we think it is  adequate for our purpose of the data modeling.      

\begin{figure}
\includegraphics[width=4.3cm,height=8.1cm,angle=-90]{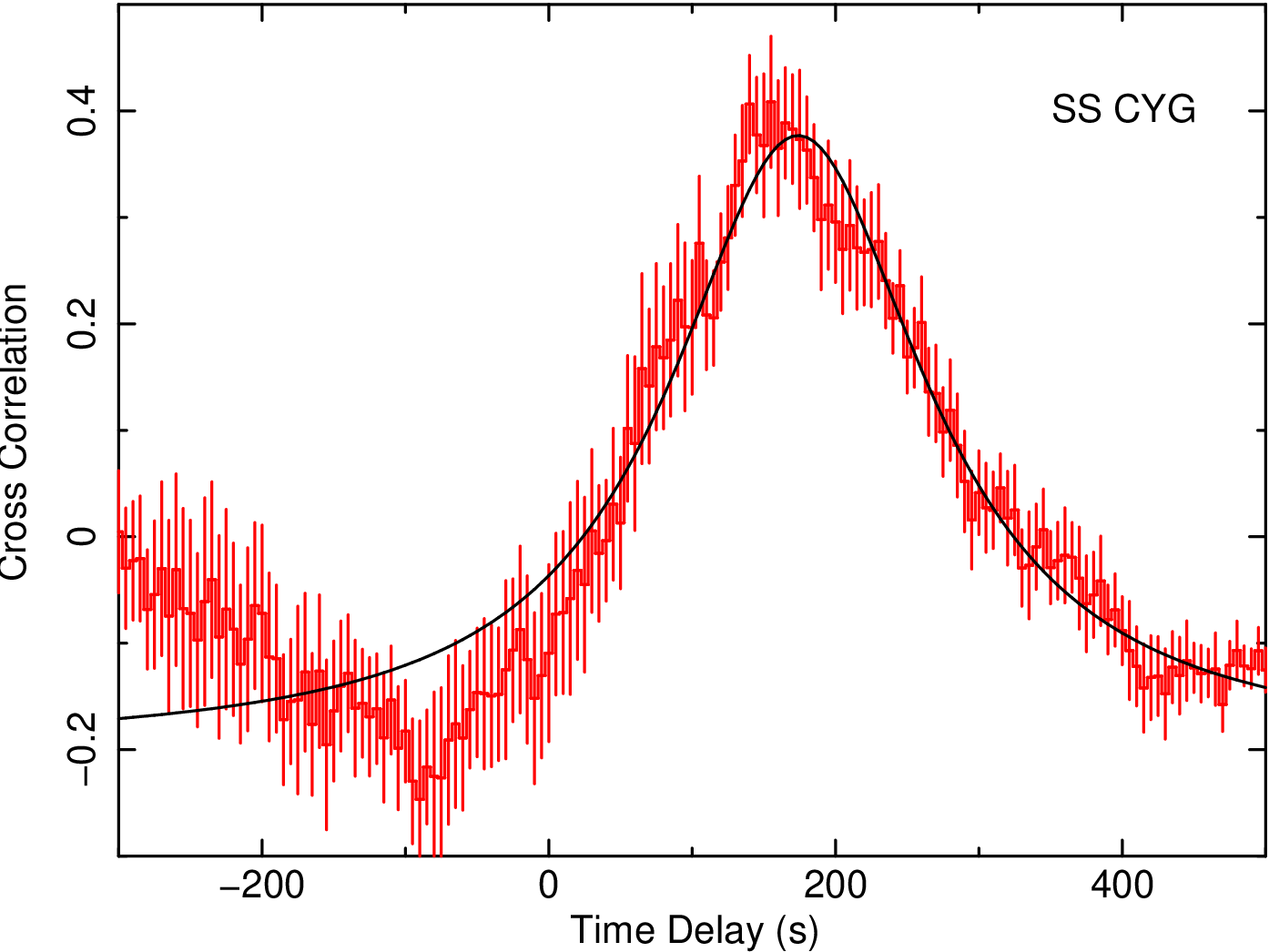}
\includegraphics[width=4.3cm,height=8.5cm,angle=-90]{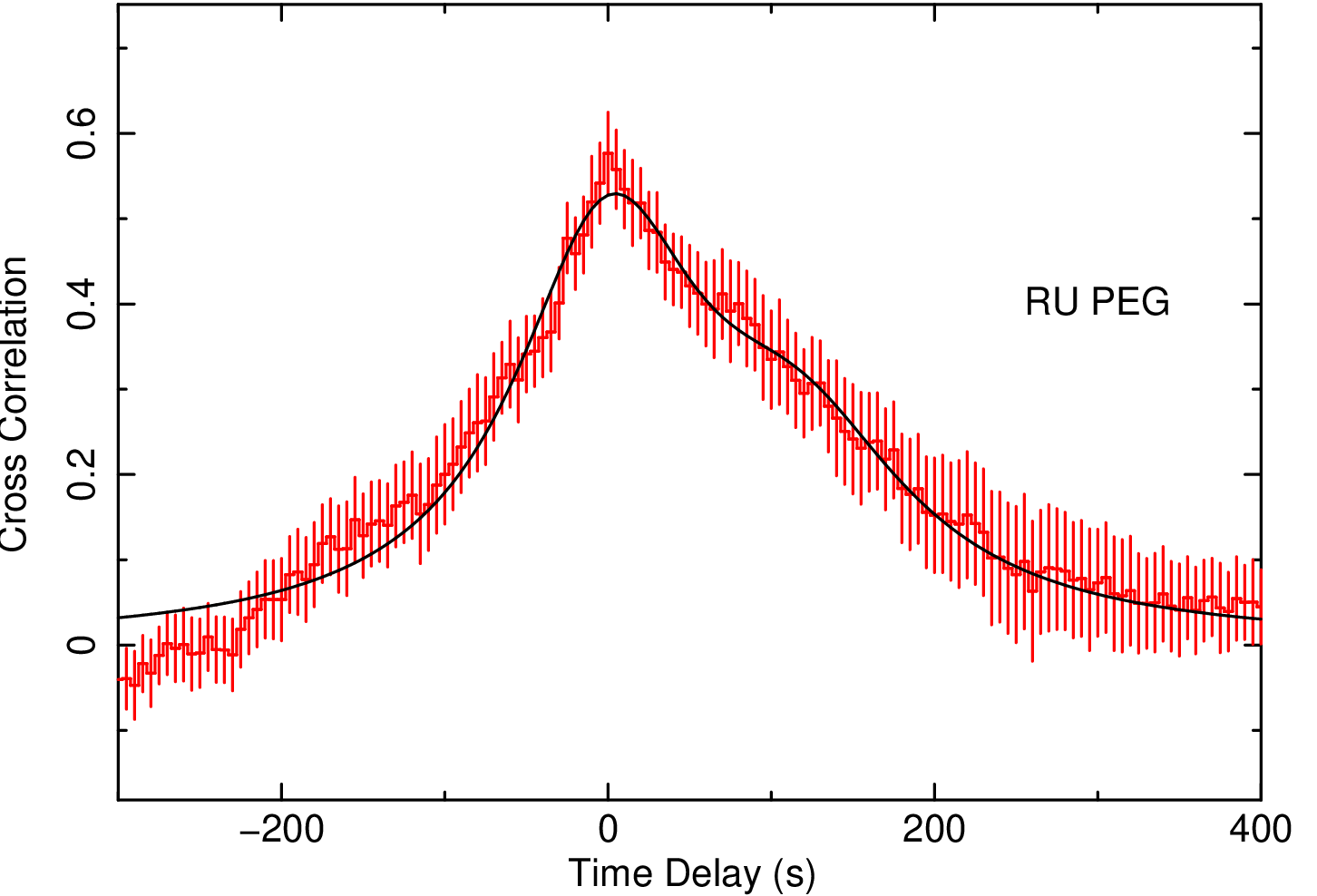}
\includegraphics[width=4.3cm,height=8.1cm,angle=-90]{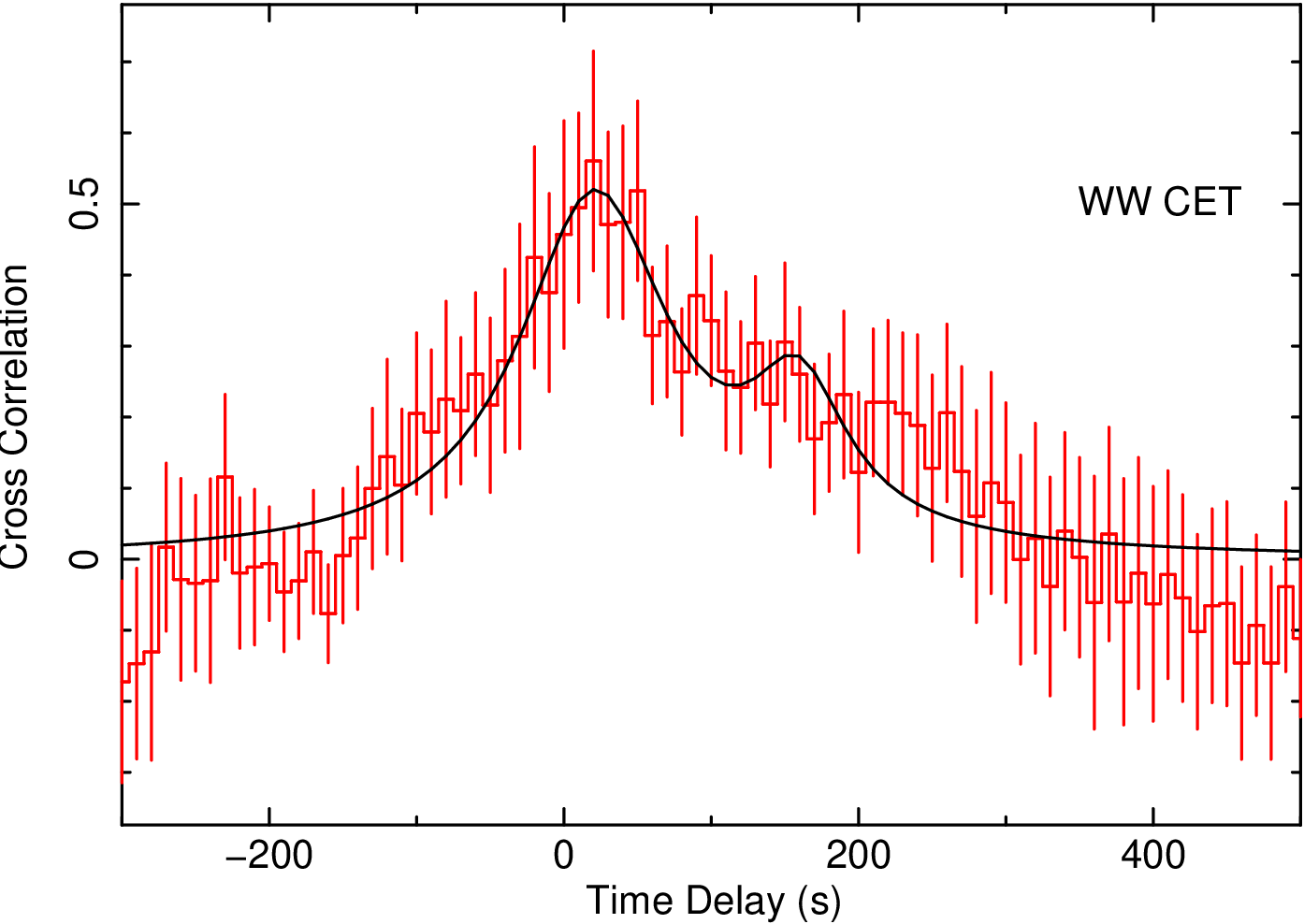}\\
\includegraphics[width=4.3cm,height=8.5cm,angle=-90]{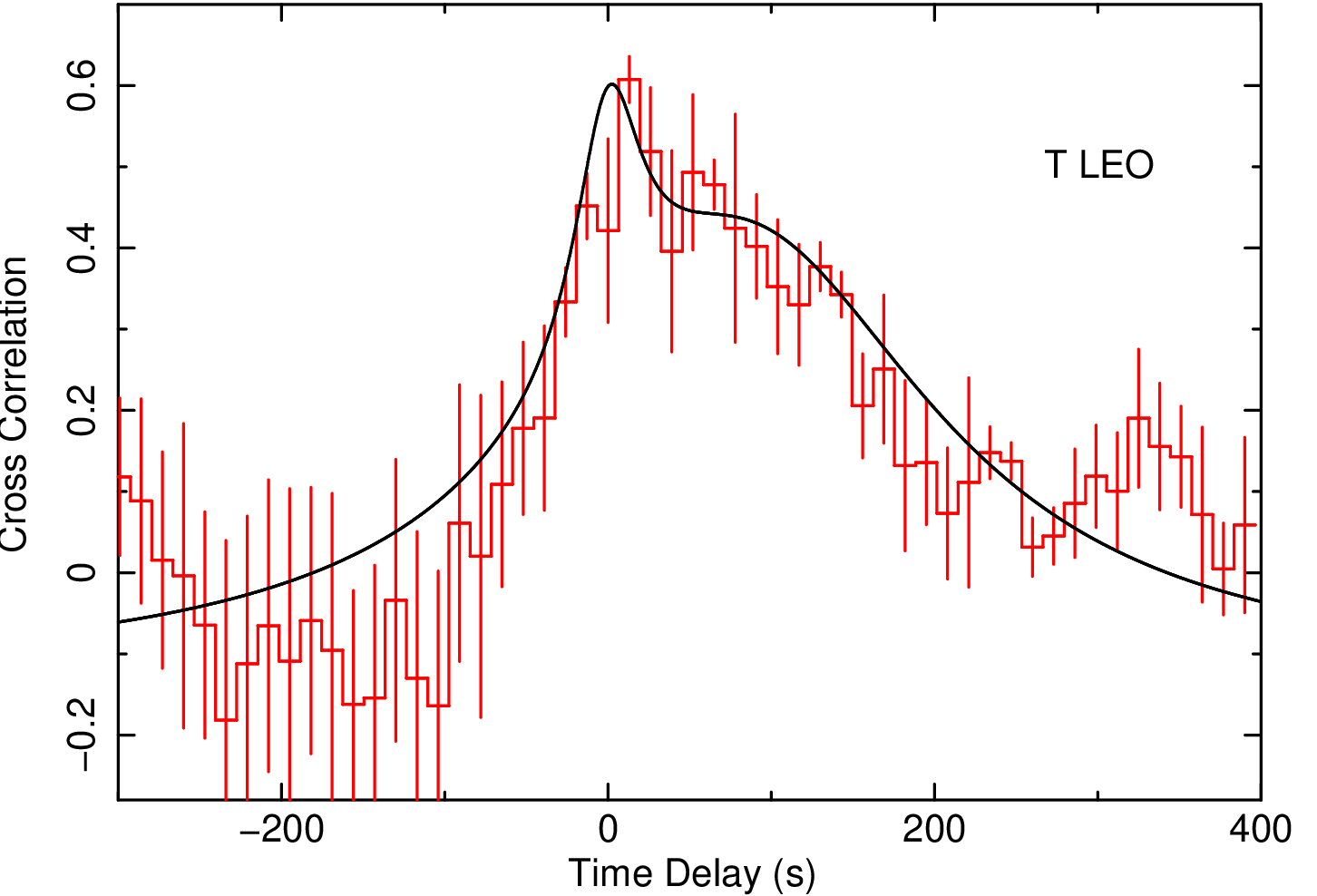}
\includegraphics[width=4.3cm,height=8.1cm,angle=-90]{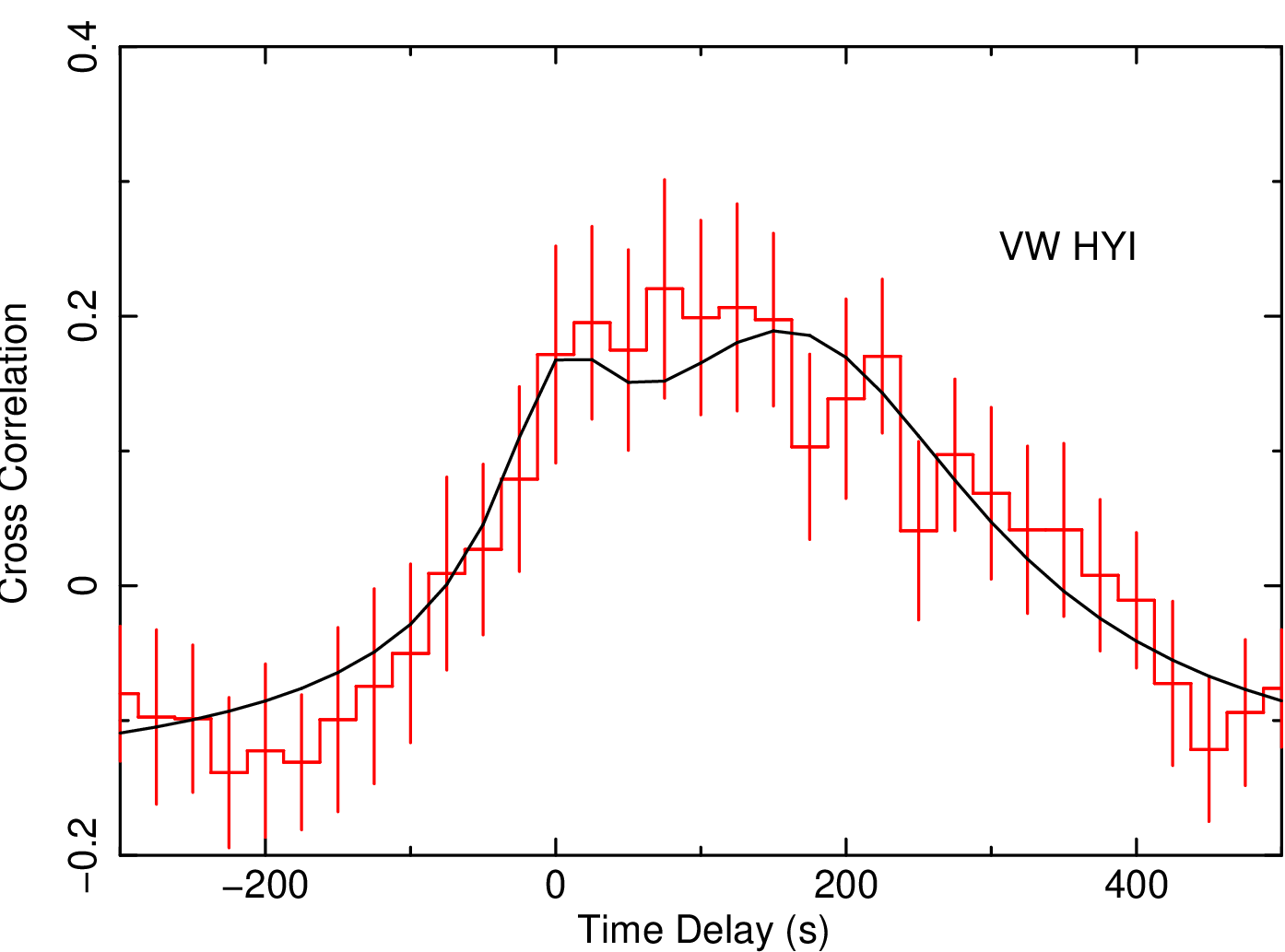}
\caption{The cross-correlation of the EPIC pn (X-ray) and OM (UV) light curves with 1 sec time resolution.  The
CCFs are displayed for SS Cyg, RU Peg, WW Cet, T Leo, and VW Hyi from the top to the bottom of the figure, respectively.
The correlation coefficient is normalized to have a maximum value of 1. The two-component Lorenzian fits 
are shown as solid black lines 
(except for SS Cyg where a single Lorentzian was used). The reduced \chisq values are 0.8, 0.4, 0.45, 1.2, and 0.45 
from the top to the bottom panel of the figure, respectively.}
\end{figure}

\begin{figure}
\includegraphics[width=5.33cm,height=7.62cm,angle=-90]{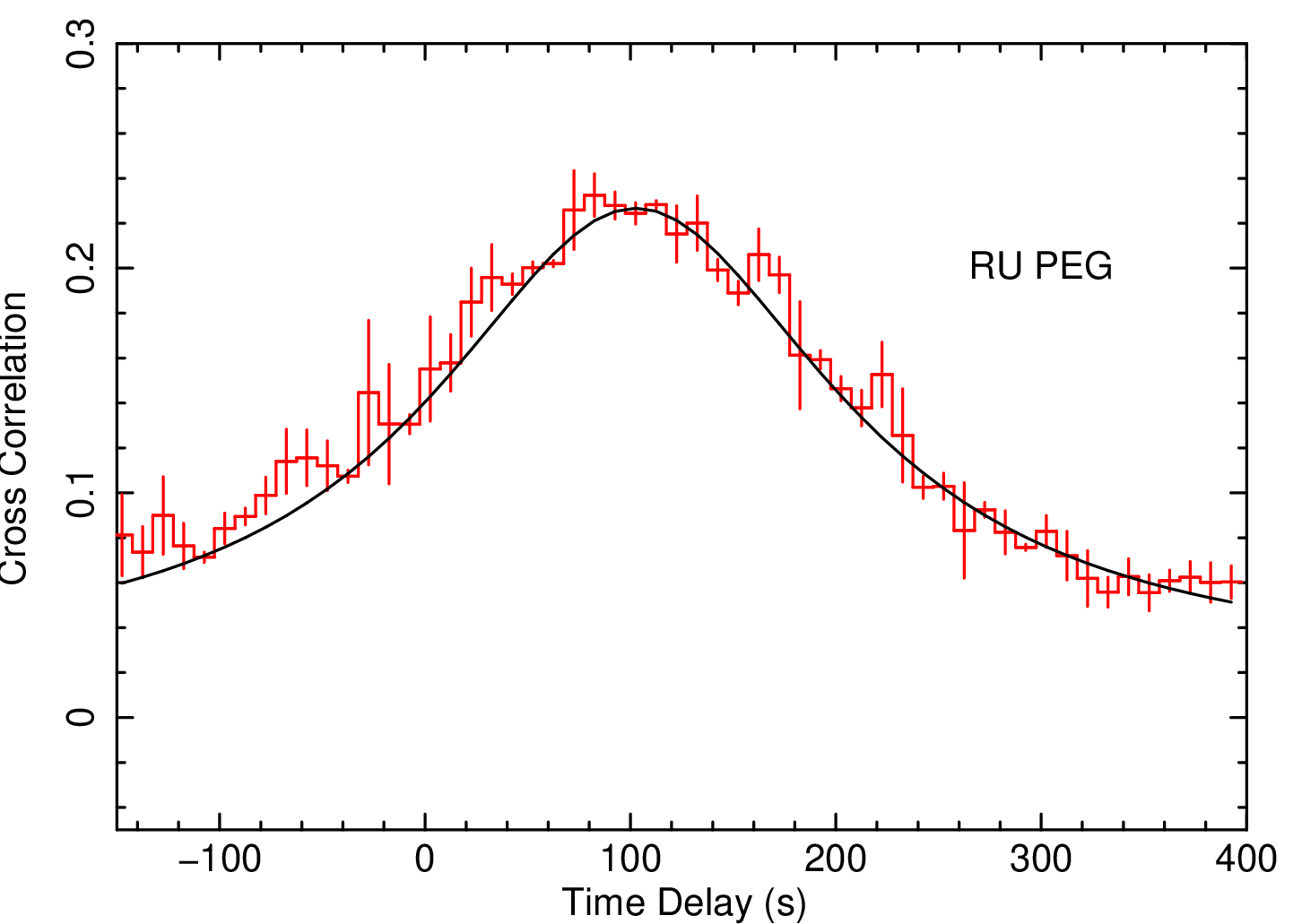}
\includegraphics[width=5.33cm,height=7.36cm,angle=-90]{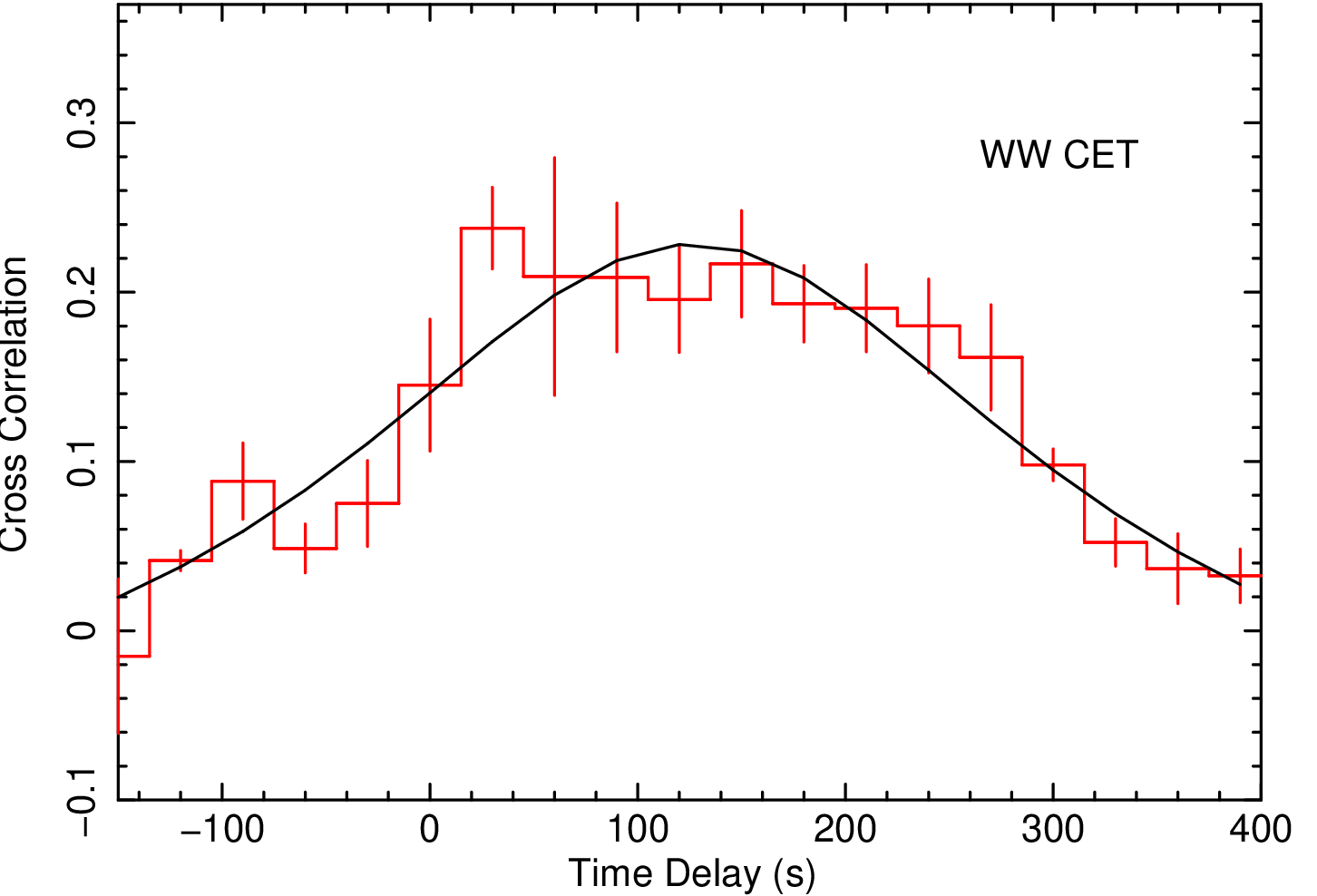}
\includegraphics[width=5.33cm,height=7.62cm,angle=-90]{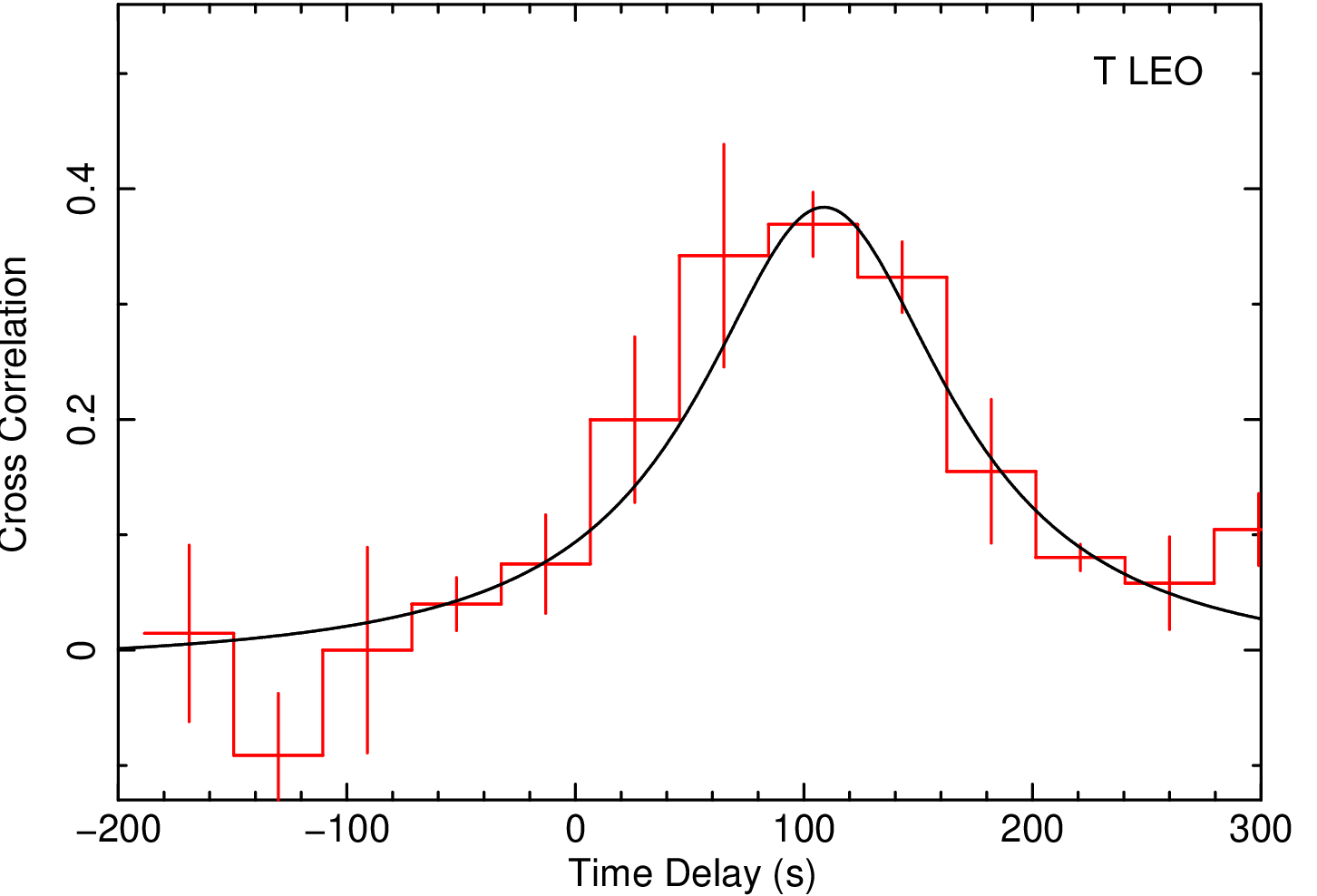}
\includegraphics[width=5.33cm,height=7.62cm,angle=-90]{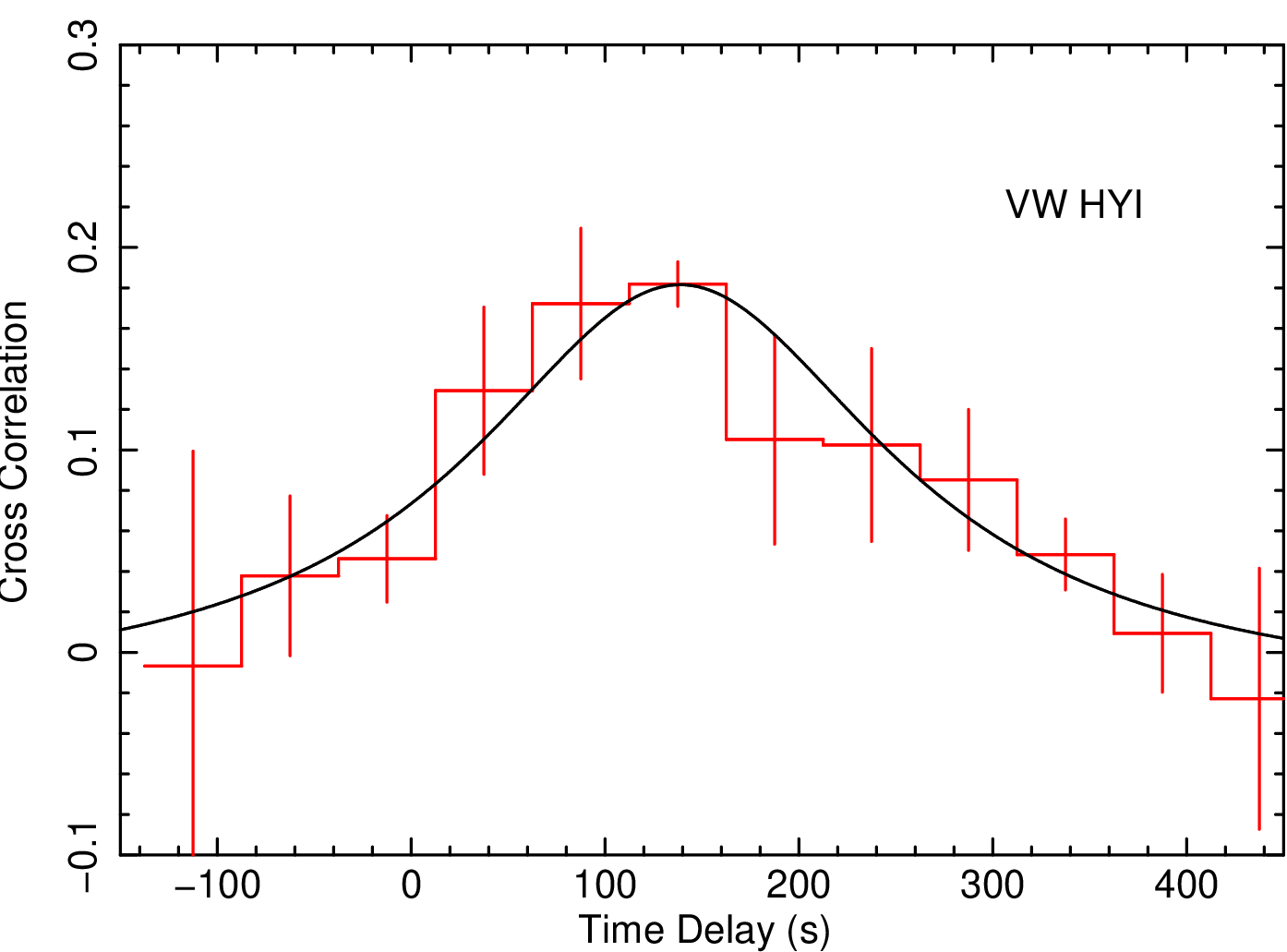}
\caption{The subtracted cross-correlation of the EPIC pn (X-ray) and 
OM (UV) light curves (see the text for details). 
The residuals of the fits to the cross-correlations 
are displayed for RU Peg, WW Cet, T Leo, and VW Hyi 
from the top to the bottom of the figure, respectively.
Single Lorenzian fits applied  are shown as solid black lines
(SS Cyg has been excluded since only a single Lorentzian was already used).  The reduced \chisq values
of the fits are given in Table 2.}
\end{figure}

\begin{table*}
\label{2}
\caption{The break frequencies, disk truncation radii, and time delays of the Dwarf Novae analysed in this work.
The last column is the logarithm of the ratio between the UV and X-ray luminosities. The UV Luminosities
are in erg s$^{-1}$ \AA$^{-1}$\ obtained using the \XMM OM UVW1 filter. The errors represent 90 $\%$ confidence level.
The \rchisq values are for the delays calculated from fits with the subtracted CCFs except for SS Cyg forwhich a single Lorentzian fit was applied. } 
\begin{center}
\begin{tabular}{cccccc}
\hline
\hline

\multicolumn{1}{c}{Source}  &
\multicolumn{1}{c}{State}  &
\multicolumn{1}{c}{Break Freq. (mHz)} &
\multicolumn{1}{c}{Radius ($\times$10$^9$ cm) } &
\multicolumn{1}{c}{Delay (s)} &
\multicolumn{1}{c}{Log(L$_{UVW1}$/L$_X$)}\\
\hline

SS Cyg &  (quiescence-XMM) & 5.6$\pm$1.4 & 4.8$\pm$1.2 & 166-181\ (\rchisq=0.8) & -4.3\\
SS Cyg &  (quiescence-RXTE) & 4.5$\pm$1.3 & 5.5$\pm$1.8 &  & \\
SS Cyg & (X-ray Dips)  & 50.0$\pm$20.0 & 1.1$\pm$0.5 & & \\
SS Cyg & (X-ray peak) & 9.7$\pm$1.5 & 3.3$\pm$0.5 &  &  \\
RU Peg & (quiescence) &2.8$\pm$0.5  & 8.2$\pm$1.5 & 97-109\ (\rchisq=1.7) & -3.1 \\
VW Hyi & (quiescence) & 2.0$\pm$0.6 & 8.1$\pm$2.5 & 103-165\ (\rchisq=0.5) & -2.7 \\
WW Cet & (quiescence) & 3.0$\pm$1.7 & 6.8$\pm$3.8 & 118-136\ (\rchisq=1.6) & -3.6 \\
T Leo & (quiescence)  & 4.5$\pm$1.5 & 4.0$\pm$1.3 & 96-121\ (\rchisq=1.4) & -3.7 \\
\hline
\end{tabular}
\end{center}
\end{table*}

\section{Disc truncation and the matter propagation}

We have done a detailed X-ray and UV power spectral analysis of SS Cyg during quiescence
and outburst together with the 
power spectral analysis of the X-ray and UV data of four other
DN systems. We have investigated the red noise structure resulting from the flicker noise
in the accretion disks using a propogating fluctuations model and fitted the PDS yielding the break frequencies where 
this noise
starts to subside. This is a strong indication that the optically thick Keplerian flow is 
truncated at some large radii during the quiescent state of DNe. Note that similar conclusions 
was recently reached via analysis of fast optical variability of SS Cyg in quiescence 
and in outburst by \cite{revnivtsev12}. 

We calculated a range of disk truncation radii (10-0.3)$\times$10$^9$ cm
with the WD masses 0.4-1.3 M$_{\odot}$. These results are  consistent with the theoretical calculations
from the irradiation of WDs \citep{king97}, and the disk evaporation and formation of coronal flows 
\citep{meyer94,liu97,mineshige98,dekool99}. WD irradiation yields a truncation radius about (1-5)R$_{WD}$; the 
disk evaporation and formation of coronal flows predict inner disk radii in a range 2.0-6.0$\times$10$^9$ cm. 
These ranges are consistent and within the errors of all the disk truncation radii we calculated in this work. 

General picture of the accretion flow around a WD in quiescence thus might be somewhat similar to that of the  
black hole/neutron star accretors with an optically thick cold outer accretion disk and an optically thin hot inner corona 
\cite[see e.g.][]{esin97}. The appearance of hot corona in the innermost regions of the flow may significantly differ from 
that of ordinary rotating Keplerian disk because it is no longer fully supported by rotation, but might have a significant 
radial velocity component. In this case the region where matter settles to the WD surface might differ from a 
simple accretion belt, anticipated by e.g., \cite{kippenhahn78}. The radial motion of the matter in the coronal flow above the disk 
(close to the truncation radius) might lead to observational appearance of single broadly peaked emission lines in 
eclipsed systems, discussed e.g. by \cite{williams89} (note, that optical emission lines might originate also not in the 
innermost regions of the flow, see e.g. \citealt{groot01}). However, existence of accretion columns on the WD surfaces 
in the case of the studied dwarf  novae in this paper is questionable because of the absence of any detectable pulsed X-ray emission.
Moreover, it has been observed that the eclipsing dwarf nova exhibiting double-peaked emission lines shows evidence
of extended emission during the eclipse with plasma temperatures different than the persistent emission e.g. \cite{ramsay01}.

The only dwarf novae, for which we could trace the X-ray PDS evolution through several outbursts with statistically good data 
was SS Cyg. We detect that the disk moves in from (6-5)$\times$10$^9$ cm to about 1.1$\times$10$^9$ cm
from the quiescence to the peak of the optical outburst. After that it eventually moves out as the 
X-rays rise and go back to quiescence (note also similar results from fast optical 
timing study of SS Cyg by \citealt{revnivtsev12}). This is a strong indication of the disk truncation  scenario and its applicability 
to the explanation of the UV/X-ray delays during the DN outbursts.

The $Suzaku$ observations
of SS Cyg during the outburst stage support the existence of a disk corona as calculated from the reflection
off of the disk using particularly the 6.4 keV iron fluorescence line \citep{ishida09}.
The authors find that the coronal region is at r$<$7$\times$10$^9$ cm which is consistent with the
disk truncation radii we calculate in this work.
The $Suzaku$ data obtained in quiescence may also be consistent with a coronal structure 
(Ishida 2011, private communication).
The combined $FUSE$ and $HST$ data analysis of SS Cyg (in the UV wavelengths) also reveal that almost all the lines
are in emission possibly from an optically thin region in the disk and/or corona making it difficult to assess
rotational broadening and chemical abundances based on absorption lines and the data are not consistent
with the  simple standard disk models \citep{sion10}.

We checked the correlation between the variability of the X-ray and the UV data using \XMM EPIC pn and OM.
We found two components in four systems; one delayed and the other undelayed (at least down to the time resolution of 
the used 
data). The only peculiarity in this respect is the case of SS Cyg (the fifth source), which shows only a delayed component. 
The significant variability correlation at $\Delta$t$\sim$0 lag is 
expected to
be caused by the reprocessing of X-rays (i.e., irradiation by X-rays) in the
accretion disk. Such time lags are on the order of milliseconds and
proportional to the light travel time which is beyond the time resolution
in our UV light curves. The delayed component detected in the five systems
with  96-181 sec lag
is much longer compared with the light travel effects. In the frame work of the
model of propagating fluctuations, this time lag should appear due to the finite time needed for matter to travel 
from the innermost parts of the accretion disk (UV emitting region) to the surface of the WD, where the majority of the 
X-ray emission is created.

A similar delay of X-rays with respect to the variations of the optical flux was detected in the
magnetic accreting WD - EX Hya (Revnivtsev et al. 2011). In this case, it is a known fact that the accretion 
disk is truncated due to the interaction with the WD magnetosphere. The time lag detected 
in this system is around 7 seconds (Revnivtsev et al. 2011), while the break frequency is 
$\sim(1.7\pm0.2)\times10^{-2}$ Hz. 

Note that the ratio of these two time scales -- time scale, defined by the break in power 
spectrum $t_{\rm break}$ and the time lag $t_{\rm lag}$ -- is significantly different between 
the case of magnetic accreting WD, EX Hya, and the case of non-magnetic, DNe. This is expected in the 
framework of the model of propagating fluctuations. 

As shown in Revnivtsev et al. (2009), 
$t_{\rm break}$ is close to the time of rotation of matter on the Keplerian orbit at the inner 
edge of the disk. 
The travel time ($\approx$ time lag) in the case of magnetic Intermediate Polar systems 
is approximately equal to the time 
of free fall from the inner boundary of the disk to the WD surface (matter moves along the magnetic 
field lines in the WD magnetosphere). 

In the case of DNe which are non-magnetic systems, this timescale is a time of 
viscous propagation of matter fluctuations from the inner radius of the optically thick accretion disk to the region  
 where the bulk of the X-ray generation occurs 
-- close to the WD surface \citep{mukai97,nucita09}. We should emphasize here that below the truncation radius the accretion 
flow is optically thin and thus -- hot, therefore its viscous properties are different from those of the outer non-active disk. 

If we assume that the innermost 
parts of the accretion flow is optically thin and virialized, then the radial velocity $v_{\rm r}$ in this inner part of the flow is $v_{\rm r}\sim \alpha v_{\rm K}$, where $v_{\rm K}$ and is 
the Keplerian velocity , while in the case of a simple free-fall, the radial velocity can be 
estimated as $v_{\rm r}\sim v_{\rm K}$. The ratio of $t_{\rm break}/t_{\rm lag}$ in these two 
cases (magnetized and non-magnetized CVs) should depend on $\alpha$.  The 
ratio of $t_{\rm break}/t_{\rm lag}$ approximately equals (as expected) to $\sim 2\pi$ in the case of a magnetic CV EX Hya \citep{revnivtsev11}, and it is approximately 4 times smaller in the case of the non-magnetic DN systems, yielding a value of $\alpha\sim 0.25$ for the hot inner part of the accretion flow. We stress that this estimate should be treated with caution because of 
our very simplified approach. Note that our viscosity estimate significantly depends on our conclusions that the disk is 
truncated at large distances ($>1-5~R_{\rm WD}$) from the WD surface. In some previous studies where the matter travel distance was 
assumed to be much smaller than we deduce from our analysis, authors obtained much smaller values of $\alpha$ in the inner 
accretion flow where they presumed a standard optically thick disk reaching to the WD with a narrow boundary layer 
bright and optically thin in the X-rays \cite[e.g.][]{godon05}.

\section{Summary and Conclusions}
 
We have presented the power spectral analysis of five DNe systems in quiescence and
searched for cross-correlations between the X-ray and UV light curves. 

We have studied the red noise structure resulting from the flicker noise
in the accretion disks and modeled the PDS yielding the break frequencies. 
This is a strong indication that the optically thick Keplerian flow is
truncated at some large radii during the quiescent state of DNe and coronal
flows are formed (optically thick-thin disk transition). These structures may be extended on the accretion disk and
be emitting at low levels that require  high sensitivity for detection.
Such structures may be created by the disk evaporation as suggested earlier.
Our range of break frequencies (1-6 mHz) yield 
a range of radii (10-0.3)$\times$10$^9$ cm
for the inner disk radius with the WD masses 0.4-1.3 M$_{\odot}$. 

In addition, we have also analysed the X-ray outburst data of SS Cyg
taken at different epochs and derived that the disk  moves in
from the large truncation radius during quiescence towards the WD during
the optical peak of the outburst and recedes as the X-rays peak in the outburst, finally to 
the quiescent inner disk radius.  Our findings are consistent with the
previous suggestions of an accretion disk corona existing in SS Cyg.

We modeled the cross-correlations of the quiescent UV and X-ray light curves of our sample of DNe
yielding time delays of X-rays in the range 96-181 sec  which indicates the time-lag of emission as the matter travels 
from the innermost parts of a truncated
accretion disk (UV emitting region) to the surface of the WD, where the majority of the 
X-ray emission is created (i.e., the X-ray photons are delayed). In four of the systems (except for SS Cyg) 
we also detect zero-time lag correlation indicating the existence of 
irradiation and reprocessing of X-rays from the cold disk consistent with the light-crossing
timescales of the systems. 

Finally, in the framework of the propagating fluctuations model we used the ratio of the break timescale and the time-lag of magnetic CV 
EX Hya and DN systems to derive an estimate of 0.25 for the $\alpha$ parameter in the inner (optically thin) parts of the accretion 
flow of DNe disks.

\section*{Acknowledgments}
SB acknowledges financial support from both 
TUBITAK National Observatory (TUG) and Space Research Institute of 
Russian Academy of Sciences for her visits to Space Research Institute (IKI)
in 2010-2011. MR acknowledge the support by the grants of President of Russian 
Federation MD-1832.2011.2, RFBR 10-02-00492, program P21 of
Presidium of the Russian Academy of Sciences/RAS, program OFN17 of the Division 
of Physical Sciences of the RAS and Dynasty Foundation.

\end{document}